%% file: p.tex
\documentclass[aps,amsmath,amssymb,superscriptaddress]{revtex4-2}

\usepackage{comment}
\usepackage[]{graphicx}
\usepackage{dcolumn}
\usepackage{bm}
\usepackage{hyperref}
\usepackage{xcolor}
\usepackage{breqn}
\graphicspath{{./}}

\newcommand{\e}{\mbox{e}}
\newcommand{\vect}[1]{\mathbf{#1}}
\newcommand{\baseU}{\overline{U}}
\newcommand{\baseW}{\overline{W}}
\newcommand{\baseP}{\overline{P}}

\newcommand{\hv}{\hat{v}}

\newcommand{\heta}{\hat{\eta}}

\newcommand{\at}{\tilde{\alpha}}
\newcommand{\p}{\partial}

\begin{document}


\title{Linear stability of Poiseuille flow over a steady spanwise Stokes layer}

\author{Daniele Massaro}
\email{dmassaro@kth.se}
\affiliation{SimEx/FLOW\char`,{} Engineering Mechanics\char`,{} KTH Royal Institute of Technology\char`,{} SE-100 44 Stockholm\char`,{} Sweden}
\affiliation{Department of Aerospace Sciences and Technologies\char`,{} Politecnico di Milano\char`,{} via La Masa 34 20156 Milano\char`,{} Italy}
\author{Fulvio Martinelli}
\affiliation{Laboratoire d'Hydrodynamique (LadHyX)\char`,{} CNRS--\'Ecole Polytecnique\char`,{} F-91128 Palaiseau\char`,{} France}
\author{Peter Schmid}
\email{peter.schmid@kaust.edu.sa}
\affiliation{Department of Mechanical Engineering\char`,{} PSE Division\char`,{} KAUST\char`,{} 23955 Thuwal\char`,{} Saudi Arabia}
\author{Maurizio Quadrio}
\email{maurizio.quadrio@polimi.it}
\affiliation{Department of Aerospace Sciences and Technologies\char`,{} Politecnico di Milano\char`,{} via La Masa 34 20156 Milano\char`,{} Italy}

\begin{abstract}
The temporal linear stability of plane Poiseuille flow modified by
spanwise forcing applied at the walls is considered. The forcing
consists of a stationary streamwise distribution of spanwise velocity
that generates a steady transversal Stokes layer, known to
reduce skin-friction drag in a turbulent flow with little energetic
cost. 
A large numerical study is carried out, where
the effects of both the physical and the discretization parameters are
thoroughly explored, for three representative subcritical values of the Reynolds
number $Re$. Results show that the spanwise Stokes layer significantly
affects the linear stability of the system. For example, at $Re=2000$
the wall forcing is found to more than double the negative real part
of the least-stable eigenvalue, and to decrease by nearly a factor of four 
the maximum transient growth of perturbation energy. These observations are
$Re$-dependent and further improve at higher $Re$. 
Comments on the physical implications of the obtained results are provided, suggesting that spanwise forcing might be effective to obtain at the same time a delayed transition to turbulence and a reduced turbulent friction. 
\end{abstract}

\maketitle

\input{intro}
\input{math}
\input{discretization}
\input{results}

\input{conclusions}

\begin{acknowledgements}
Preliminary results of this research were presented 
\citep{martinelli-quadrio-schmid-2011} at the XIII EUROMECH Turbulence
Conference in Warsaw. Dr. Carlo Sovardi is thanked for his support work during his Master thesis.
\end{acknowledgements}

\section*{Conflict of interest}
The authors report no conflict of interest.

\appendix
\section{Discretization parameters}
\label{sec:appendix}

For each set of physical parameters, the discretization parameters
must be chosen. Some are simply set beforehand after a preliminary
study, while others are dynamically adapted during the calculation 
on a case-by-case basis to
consistently satisfy certain criteria across the whole study. In the
following, we describe how the discretization parameters
have been chosen to ensure consistent, high-fidelity solutions. 

\subsection{Number of Chebyshev polynomials}

The number $N$ of Chebyshev polynomials used to discretize the
wall-normal direction must be sufficiently large to provide accurate
values of the most unstable eigenvalues and of the maximum $G_{max}$ of the
function $G(t).$ For standard Poiseuille flow, ample literature is
available (see e.g.\ Refs.\ \cite{orszag-1971, schmid-henningson-2001,
  reddy-schmid-henningson-1993, harfash-2015}) to inform the choice of
a suitable $N$; here, however, the presence of an additional spanwise
base flow calls for a further systematic analysis.

\begin{table}
\begin{center}
\def~{\hphantom{0}}
\begin{tabular}{c|cccc|rrrr} 
		$\kappa $ & $N=40$ & $N=80$ & $N=100$ & $N=120$ & $N=40$ & $N=80$ & $N=100$ & $N=120$ \\
		\hline
		0.5 & -0.71597 & -0.71589 & -0.71589 & -0.71589  & 89.2768 & 221.0830 & 236.2375 & 240.5213  \\
		0.75 & -0.69655 & -0.69662 & -0.69662 & -0.69662 & 56.6170 & 78.4394 & 80.0837 & 80.2476 \\
		  1 & -0.68968 & -0.68974 & -0.68974 & -0.68974 & 53.1008 & 53.2548 & 53.2840 & 53.2953 \\
		1.5 & -0.68762 & -0.68745 & -0.68745 & -0.68745  & 63.3796 & 63.4602 & 63.4707 & 63.4793\\
		2.5 & -0.75017 & -0.75032 & -0.75032 & -0.75032  & 67.9628 & 67.9501 & 67.9571 & 67.9720  \\
		3.5 &-0.80562 & -0.80556 & -0.80556 & -0.80556  &66.5897 & 66.6047 & 66.6061 & 66.6080  \\
		5 &-0.83934 & -0.83936 & -0.83936 & -0.83936  &65.7172 & 65.6933 & 65.6931 & 65.6934  \\
\end{tabular}
  \caption{Modal and non-modal stability results as a function of the number $N$ of
    Chebyshev polynomials and of the forcing wavenumber $\kappa$, for
    $Re=1000$, $\beta=1.5$ and $A=1$. Left: real part
    of the least stable eigenvalue $\lambda_1$ multiplied by $10^2$. Right: maximum value $G_{max}$ of the transient energy growth function.}
  \label{tab:N}
  \end{center}
\end{table}

A preliminary resolution study is thus carried out for a typical case,
with $Re=1000$, $\beta=1.5$ and $A=1$, where the wavenumber is varied
from $\kappa=0.5$ to $\kappa=5$. The effect of changing $N$ is
observed on the main modal and non-modal stability characteristics:
table \ref{tab:N} reports on the left the real part of the least stable
eigenvalue $\lambda_1$, and on the right the
maximum value $G_{max}$ of the transient growth function. $N$ is
varied from $N=40$ to $N=120$. The modal stability characteristics are
weakly sensitive to the value of $N$, with $N = 80$ already providing
the first eigenvalue accurate up five digits regardless of the value
of $\kappa$. The non-modal results, however, show a strong dependence
on $N$, especially for the lowest values of $\kappa$. When $\kappa
\geq 1$ the effects are minor, and $N=80$ provides $G_{max}$-values
that are stable to the fourth digit or higher. At lower $\kappa$, the
accuracy degrades, and the largest value of $N=120$ still does not
provide resolution-independent results for $\kappa=0.5$. Fortunately,
as will be shown below, this case at the lowest $\kappa$ shows changes
in $G_{max}$ of less than 2\% when moving from $N=100$ to $N=120$ and
is of no practical interest. Moreover, the subsequent case with
$\kappa=0.75$ displays far lower sensitivity to $N$, reduced by one
order of magnitude.

The above considerations lead to our operational choice of using
$N=80$ Chebyshev polynomials when $\kappa \ge 1$, but increasing this
number to $N=100$ when $\kappa <1$. The selected resolution is further
checked on the most demanding situation, i.e. non-modal stability
calculations at the highest $Re$ considered in the present work
($Re=2000$). For $N=100$ polynomials, we compute a maximum transient
growth at $\kappa=0.75$ of $190.8346$, which is to be compared with a
value of $190.8451$ obtained with $N=120$ polynomials.


\subsection{Truncation factor}
\label{sec:Mpar}

\begin{figure}
\centering
\includegraphics[trim=5cm 3.5cm 4.8cm 2.5cm,clip,
                 width=0.9\textwidth]{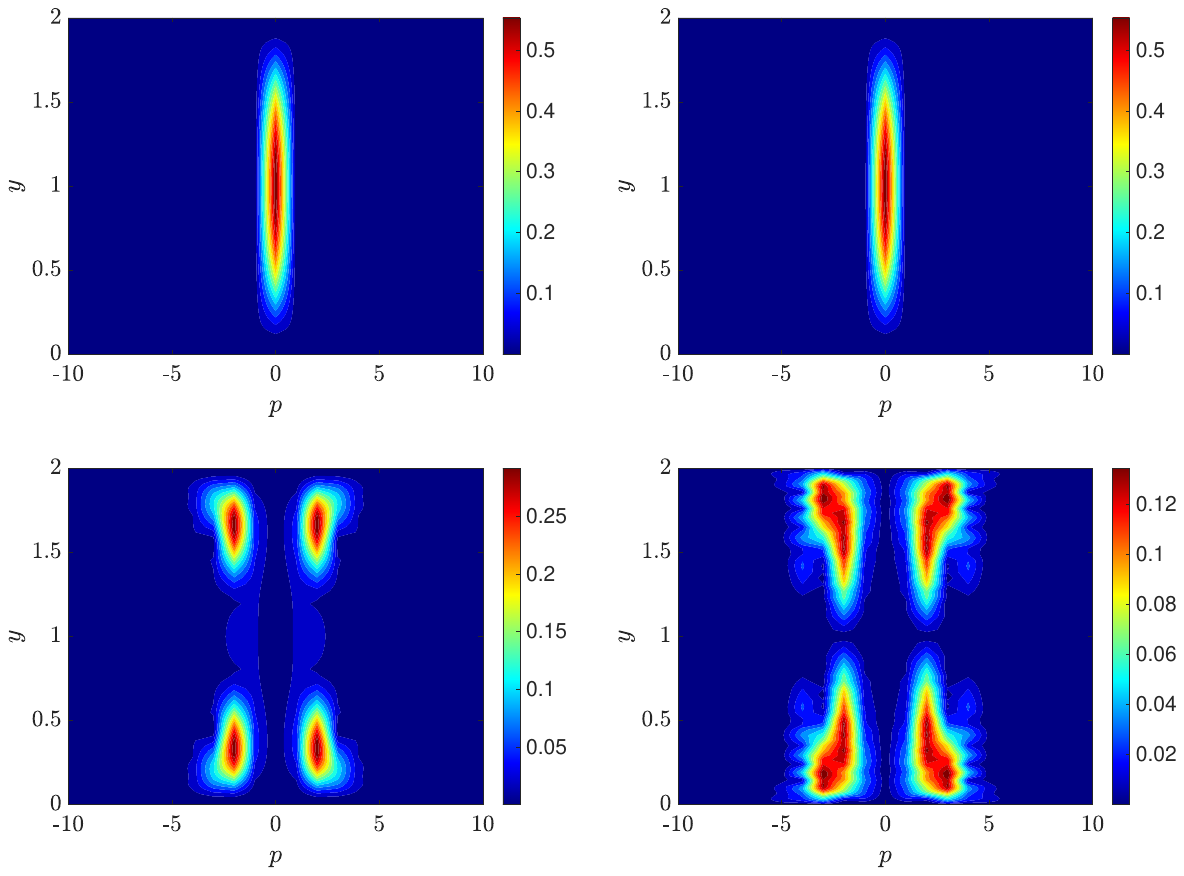}
\caption{Optimal initial condition, at $Re=1000$ for $v$ (left) and $\eta$ (right)
  as a function of the $y$ coordinate and the $p$-th wavenumber of the
  modal expansions, $-M \le p \le M$. Contours describe the absolute
  value of the optimal initial condition. Top: example of optimal
  input centered around $p=0$, with $\kappa=1.5$, $A=0.3$
  and $\beta=0.5.$ Bottom: example of optimal input peaking at $\alpha
  \sim \pm \kappa^{-1}$ with $\kappa=1$, $A=0.6$ and
  $\beta=0.3.$ }
\label{fig:CompactSupport}
\end{figure}

The truncation factor $M$ in Eq.\eqref{eq:xFT} defines the spectral
expansion of the flow variables in the streamwise direction. The
computed values of both $\lambda_1$ and $G_{max}$ should be robust to
the chosen value of $M$ --- a parameter which also strongly affects the
computational cost.


Since in non-modal stability analyses the optimal perturbation is not
constrained to a single streamwise wavenumber, $M$ should be
sufficiently large to guarantee a compact support for the optimal
initial condition, while also accurately capturing its evolution up to
the time of maximum energy growth. We have empirically established
that the optimal initial conditions obtained by varying the problem
parameters can be grouped into two broad classes with qualitatively
different characteristics in terms of spectral content. Figure
\ref{fig:CompactSupport} shows, for the two classes of optimal initial
conditions, how their amplitude depends on the wall-normal coordinate
$y$ and on the modal expansion index $p$. In the first class (top row
of figure \ref{fig:CompactSupport}), the $v$-component is largest at
$\alpha=0$, i.e. $p=0$, whereas the $\eta$-component is largest near
the walls at $\alpha \sim A/\kappa$. In the second class (bottom row),
both $v$ and $\eta$ show maxima at a wavenumber inversely proportional
to $\kappa$. The temporal evolutions of perturbations from either class
are qualitatively similar.

Owing to the procedural complexity of properly selecting the parameter $M$, our operational choice is to enforce a dynamical and automated adjustment of $M$. Each simulation is started with a reasonable first guess $M_0$, and then $M$ is incremented by unitary steps until a most unstable eigenvalue $\lambda_1^M$ is computed that satisfies the criterion
\begin{equation}
\left| \frac{\lambda^M_1 - \lambda^{M-1}_1 }{\lambda_1^M} \right| < 1
\times 10^{-6}.
\label{eq:iteration}
\end{equation}

The initial guess $M_0$ is selected by empirically accounting for the
two distributions discussed above, according to the following
heuristics that depends on the forcing parameters:
\begin{equation}
  M_0 = 3 \frac{A}{\kappa} + \frac{1}{\kappa}.
\label{eq:M0}
\end{equation}
Once $M_0$ is set, an iterative increase of $M$ is started until the
termination criterion \eqref{eq:iteration} is satisfied (or a maximum
value of $M_{max}=50$ is reached). Across the entire study, this
iteration always terminated before the limit $M=M_{max}$ is reached,
and the largest value used within our study was $M=18$.

\begin{table}
  \centering
  \begin{tabular}{ccccccc} 
    $M$ &  & $8$    &  $10$  & $11$   & $14$   & $18$ \\ 
    $G_{max}$     &  & 231.1303 & 231.1788 & 231.2086 & 231.2300 & 231.2390  \\
  \end{tabular}
  \caption{Non-modal stability results as a function of the modal
    truncation factor $M$ for $Re=2000$, $\kappa=0.75$, $\beta=2$ and
    $A=1$.}
  \label{tab:Mval}
\end{table}

Provided that $M$ is not too low, the values of $G_{max}$ are not
particularly sensitive to $M$, as shown in table \ref{tab:Mval}. For
this case, our procedure leads to $M=11$, and the relative change of
$G_{max}$ for value of $M$ between $M=11$ and $M=18$ is bounded by
$10^{-3}$.

\subsection{Number of eigenvalues}

The fraction of the entire set of $N_{tot} = 2(N + 1)(2M + 1)$
eigenvalues of the system matrix retained in the iterative Arnoldi
procedure must ensure that the dynamics of the optimal initial
condition is well represented. The number $n_{eig}$ of retained
eigenvalues defines the truncation operator $\textbf{T}$, which
reduces the size of the matrix $\textbf{Q}_s$ and with it the
cost of computing the transient growth rate
\eqref{eq:gFunc}.

The worst-case scenario corresponds to optimal initial conditions of
the type illustrated in the bottom row of Figure
\ref{fig:CompactSupport}. For one such case (namely $Re=1000$, $A=0.6$
and $\beta=0.3$), table \ref{tab:nEIG} reports the computed value of
$G_{max}$ as $n_{eig}$ is changed, for selected values of
$\kappa$. The sensitivity is certainly non-negligible, especially at
the lower $\kappa$. By examining this case as well as other
representative cases, we arrive at the operational choice of setting
$n_{eig}=N_{tot}/6$ when $ \kappa \ge 1$, and $n_{eig}=N_{tot}/3$
otherwise.


\begin{table}
\centering
\begin{tabular}{cccccccc} 
 $\kappa \backslash n_{eig}$ & & $N_{tot}/7$ & $N_{tot}/6$ & $N_{tot}/5$ & $N_{tot}/4$ & $N_{tot}/3$ & $N_{tot}/2$ \\ 
     & & & & & \\
 0.5 & & 20.2629 &   20.7811  & 20.8603  & 20.8655 &  20.8657  & 20.8657 \\
0.75 &  & 15.0971  & 15.1162  & 15.1261 &  15.1266  & 15.1266  & 15.1266 \\
  1  & & 13.1278 &  13.1681  & 13.1957 &  13.2004  & 13.2013  & 13.2015 \\
 1.5 & &12.4396 &  12.4427 &  12.4431 &  12.4432 &  12.4433  & 12.4437 \\
 2.5 & &10.0582  & 10.0583 &  10.0584 &  10.0585  & 10.0588 &  10.0590 \\
 3.5 & &10.4932 &  10.4934  & 10.4937  & 10.4938  & 10.4941 &  10.4943 \\
  5  & & 10.9079  & 10.9081 &  10.9084   & 10.9084  & 10.9087 &  10.9089 \\
\end{tabular}
\caption{Non-modal stability results as a function of the number
  $n_{eig}$ of eigenvalues retained in the Arnoldi procedure and of
  the forcing wavenumber. The table reports the maximum value of the
  transient energy growth function $G(t)$.}
\label{tab:nEIG}
\end{table}

\subsection{Temporal discretization}


A temporal integration step $\Delta t$ must be chosen to evaluate
\eqref{eq:gFunc} and to identify the maximum $G_{max}$ and the time
$t_{max}$ at which it occurs. In classic, unforced Poiseuille flow,
the transient growth function $G(t)$ monotonically increases from a
unit value at $t=0$ (as long as $Re$ is above the critical value for
monotonic stability) up to $G_{max}$, and then steadily
decreases. However, when the flow is modified by the SSL, the ensuing
$G(t)$ presents two local maxima: one is similar to the unforced
maximum, the other is directly linked to the SSL. The latter is
centered on a considerably narrow peak of the $G(t)$ function, and
typically occurs at shorter times. The absolute maximum $G_{max}$ can
pertain to either peak, depending on the specific case.


\begin{figure}
  \centering \includegraphics[trim=4.5cm 0.8cm 4cm
    1cm,clip,width=0.8\textwidth]{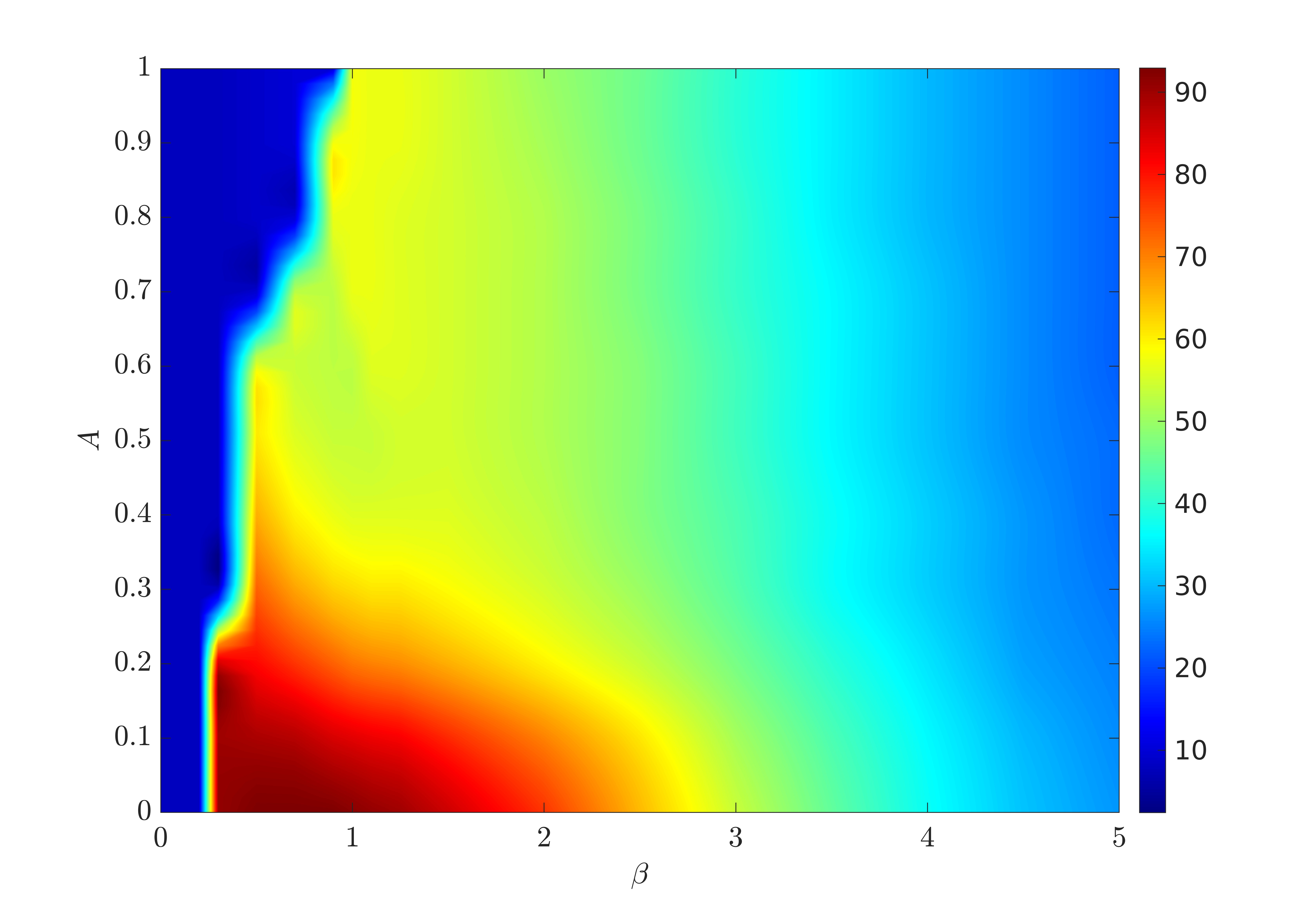}
  \caption{Time $t_{max}$ for the occurrence of the maximum $G_{max}$
    of transient growth for the case of $Re=1000$ and $\kappa=1$.}
\label{fig:tGmax}
\end{figure}

The location of the two peaks varies significantly with the physical
parameters of the problem. Figure \ref{fig:tGmax} shows, for a typical
case with dominating first peak ($Re=1000$ and $\kappa=1$), the
temporal position of the peak as a function of $A$ and $\beta$. It is
seen that, especially at small $\beta$, $G_{max}$ occurs at very short
times, whereas for small $A$ the time scale of the maximum increases,
and the unforced peak is recovered. It is also noticed that the first
peak of $G(t)$ is always located at $t<100$. Hence our operational
choice is to employ a variable time discretization by setting $\Delta
t=1$ for $0<t<100$ and increasing it to $\Delta t=10$ for $t>100$.

\bibliographystyle{unsrtdin}     
\bibliography{../../Wallturb.bib}   

\end{document}

%% file: intro.tex
\section{Introduction}
\label{sec:intro}

Decreasing the aerodynamic drag is a formidable scientific and technological
challenge in configurations dominated by a relative motion
between a solid body and a surrounding fluid. In particular, the
skin-friction drag --- in the laminar or in the turbulent regime ---
often represents a major portion of the total drag in the air transport
sector, and can be of paramount importance in naval and submarine
transport. Skin friction can be reduced either by keeping the flow laminar as
long as possible, thus exploiting the intrinsically lower friction
levels typical of the laminar regime, or by accepting the transition to
turbulence, and reducing the level of turbulent friction below its natural level. A
viable flow control approach that achieves both objectives would be
very desirable, as it would first take advantage of laminarity as long as
possible, and then continue to reduce turbulent friction.

Among the several active techniques for the reduction of turbulent
drag, we are interested in spanwise-forcing techniques, and the
present work in particular considers streamwise-traveling waves of
spanwise wall-velocity as introduced by Quadrio \textit{et al.\ }(2009)
\cite{quadrio-ricco-viotti-2009}. 
Comprehensive reviews on this technique for turbulent drag reduction are available
\citep{quadrio-2011, leschziner-2020, ricco-skote-leschziner-2021}. 
The streamwise-traveling waves, which
include as a special case the spanwise-oscillating wall
\citep{quadrio-ricco-2004} but achieve far higher energetic
efficiency, abate the levels of
turbulent skin-friction drag with interesting energetic effectiveness,
with one energy unit spent on the control saving up to 30 units of
pumping energy. Furthermore, recent evidence has shown that spanwise
forcing can bring indirect benefits in terms of the reduction of pressure
drag \citep{banchetti-luchini-quadrio-2020}; in addition, it may be
highly beneficial by interacting with the shock wave over an aerofoil in
transonic flow \citep{quadrio-etal-2022} and reducing significantly the
aerodynamic drag of the entire airplane with negligible energy expenditure. 
Within this context, the
present work is motivated by the following simple question: Can
spanwise forcing favorably affect transition to turbulence? Since the forcing is
known to weaken near-wall streaks in a turbulent flow
\citep{yakeno-2021}, a similar effect on laminar streaks might alter
their growth, thus causing a delay of, or perhaps preventing altogether,
transition to turbulence. It must be kept in mind that, at the moment,
satisfactory actuators for implementing traveling waves in a
real-world application are still lacking, even though some interesting
developments exist, including mechanical movement of the wall
\citep{auteri-etal-2010, marusic-etal-2021}, electroactive polymers
\citep{gouder-potter-morrison-2013, gatti-etal-2015} and the use of
Kagome lattices \citep{bird-santer-morrison-2018}. However, the
prospect of instrumenting e.g. an airplane wing with one actuator
that, in the wing fore part, would delay transition while, in the aft,
would decrease turbulent skin-friction drag is certainly appealing,
and motivates further research efforts into this
direction.

This work is not the first to investigate the stability properties of
a wall-bounded flow modified by spanwise forcing, and the available
body of literature provides important guidance. 
Most of the current knowledge concerns spatially uniform wall
oscillations. Jovanović (2008) \cite{jovanovic-2008}
demonstrated the capability of properly designed wall oscillations to
reduce receptivity of the linearized Navier--Stokes equations to small
stochastic disturbances in laminar Poiseuille flow. Ricco (2011)\cite{ricco-2011}
showed in a linearized study that a substantial reduction in the
intensity of laminar streaks under steady spatial oscillations is
possible, with reductions up to 90\% of the peak value of velocity
fluctuations. Rabin \textit{et al.\ }(2014) \cite{rabin-caulfield-kerswell-2014} studied plane
Couette flow under spatially-uniform wall oscillations: by solving a
fully non-linear problem, they demonstrated and quantified how the
critical disturbance energy required for the onset of turbulence
increases due to spanwise forcing. Hack \textit{et al.\ }(2012) \cite{hack-zaki-2012}, using a
linearized analysis of temporal harmonic wall oscillations, provided
further evidence for near-wall shear filtering as an effective
tool. In a follow-up study \cite{hack-zaki-2014}, the
influence of spanwise forcing on by-pass transition in the boundary
layer over a flat plate was examined. They found that oscillations, when properly
tuned, can substantially delay transition, with overall energy
gains. This DNS-based study was later corroborated by a corresponding
Floquet stability analysis \citep{hack-zaki-2015}, where the changes
in the linear modal and non-modal instability mechanisms operating in
the pre-transitional boundary layer induced by the spanwise forcing
were studied. This investigation confirmed important stabilization
mechanisms due to weaker non-modal growth, but found that transition
is enhanced, owing to a reinforcement of modal instabilities, at larger
forcing amplitudes. Similarly, Wang \textit{et al.\ }(2019) \cite{wang-liu-2019} found that
spanwise oscillations of the wall can act as precursors to the
transition process in a boundary layer.

For the spatially-varying case, where no stability study
is presently available, Duque-Daza \textit{et al.\ }(2012) \cite{duque-etal-2012} numerically solved a
linearized version of the Navier--Stokes equations for a plane channel
setting to investigate how streamwise-traveling waves impact the
growth of near-wall low-speed streaks. They found that the computed
relative change in streak amplification due to traveling waves varies
with the parameters defining the wave in a way that strictly resembles
the DNS-measured drag reduction data in the turbulent regime.
Negi \textit{et al.\ }(2015) \cite{negi-mishra-skote-2015} simulated via DNS a single low-speed
streak forced by a standing wave of spanwise wall forcing in the
laminar regime, and found that the skin friction can drop below the
laminar reference value. They also reported a delay in the
characteristic rise of skin friction during
transition. Negi \textit{et al.\ }(2019) \cite{negi-etal-2019} used LES to study bypass transition
in a spatially evolving boundary layer. Temporal and spatial
oscillations of the wall forcing were considered, and transition delay
could be observed in both cases. A qualitative explanation was
offered, attributing the transition delay to the additional filtering
of the disturbance by the Stokes layer. An optimum forcing amplitude
for transition delay was identified, while acknowledging that at
larger wavelength the delay appears to increase monotonically with
amplitude.
 

Motivated by the above studies, this work focuses on plane channel
flow as a model problem and studies the modal and non-modal stability
of the laminar, pressure-driven Poiseuille flow modified by
streamwise-distributed spanwise forcing in the form of a steady Stokes
layer, or SSL \citep{viotti-quadrio-luchini-2009}. The SSL is created
by a (spanwise-uniform) stationary spanwise velocity at the wall that
is sinusoidally distributed along the streamwise direction. The
streamwise-varying base flow prohibits the direct application of a
classic Orr--Sommerfeld--Squire (OSS) linear stability analysis, since
the resulting system of differential equations contains
streamwise-varying coefficients and thus requires a global approach. 
Here, extending an approach introduced by Floryan \cite{floryan-1997, szumbarsky-floryan-2006}, we exploit the particular form of the base flow to avoid a global stability analysis. 
The study is conceptually close to a classical secondary instability analysis, in which finite amplitude disturbances saturate and establish a new base flow, whose linear stability is then studied \citep{schmid-henningson-2001}; in the present case, it is the  superimposed sinusoidal spanwise flow that alters the Poiseuille streamwise base flow.

The structure of this paper is as follows. In \S\ref{sec:math} the
mathematical formulation is presented, emphasizing analogies and
differences with respect to a classic OSS analysis; the numerical
tools are validated against direct numerical simulations. Next in
\S\ref{sec:discretization} the physical parameters of the problem are discussed (the discussion of the discretization-related parameters is deferred to Appendix \ref{sec:appendix}), and the
computational procedures employed in the execution of a large parameter
study are described. In \S\ref{sec:results} the main results of the
modal and non-modal stability analysis are discussed, and
concluding remarks are offered in \S\ref{sec:conclusions}, together
with a critical discussion of the main findings.

%% file: math.tex
\section{Mathematical formulation}
\label{sec:math}

This section describes the setup of the linear stability analysis,
based on the well-known Orr--Sommerfeld--Squire (OSS) problem
for Poiseuille flow augmented by terms that model the presence of a
spanwise base flow induced by the forcing. 

\begin{figure}
\centering
\includegraphics[trim=4cm 3cm 3.5cm
  3cm,clip,width=0.6\textwidth]{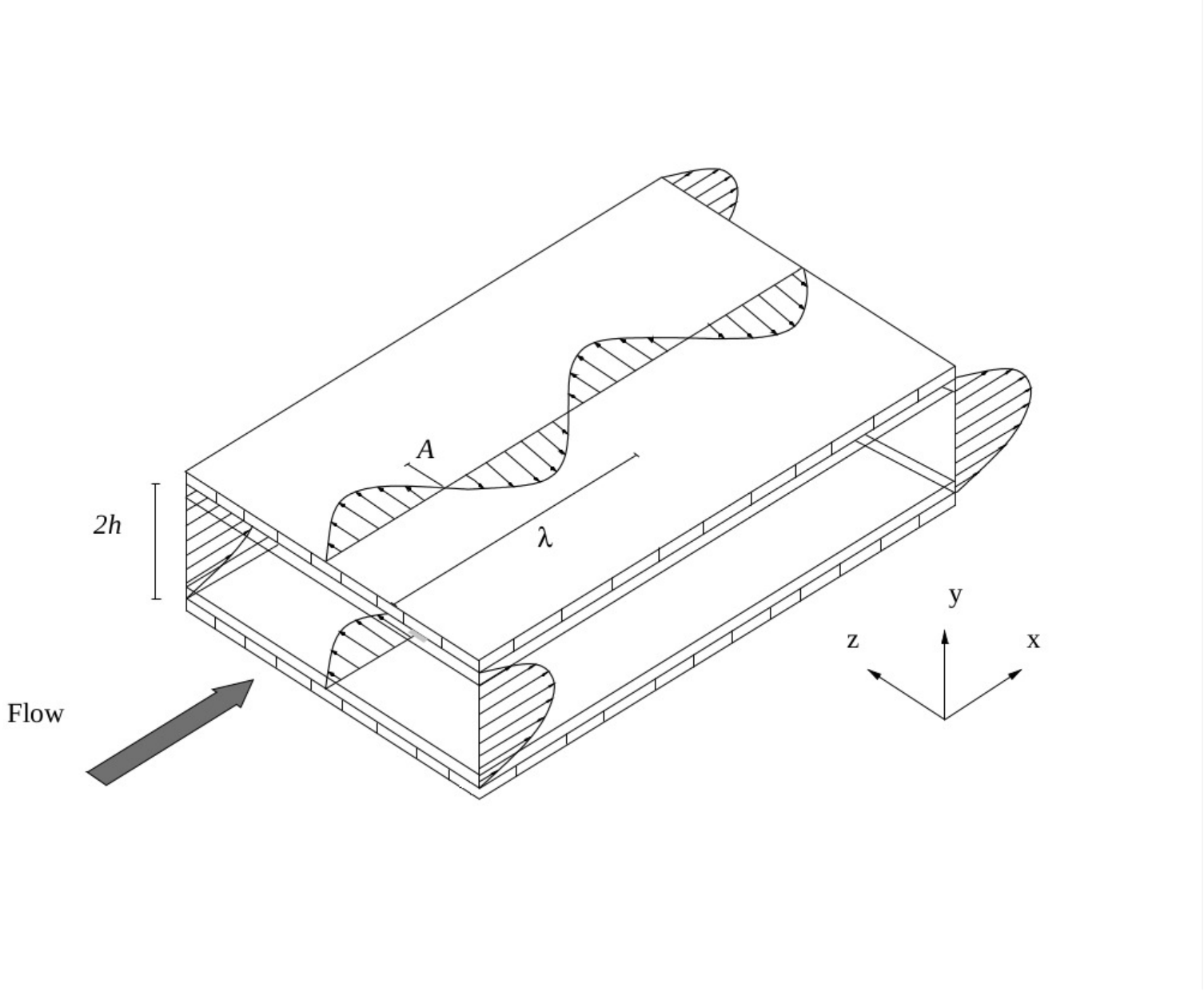}
\caption{Sketch of plane channel with the chosen reference system. The
  two channel walls are separated by a distance $2h$; a stationary
  wall forcing is applied at both walls, and is characterized by its
  wavelength $\lambda$ and maximum velocity amplitude $A$.}
\label{fig:channel}
\end{figure}

The governing equations are the incompressible Navier--Stokes (NS)
equations, which in non--dimensional form read:

\begin{equation}
  \left\{
  \begin{array}{ll}
    \nabla \cdot \vect{V} = 0 \\
    \displaystyle \frac{\partial \vect{V}}{\partial t} + \left(
    \vect{V} \cdot \nabla \right) \vect{V} = - \nabla P + \frac{1}{Re}
    \nabla^2 \vect{V}
  \end{array}
  \right.
\label{eq:ns}
\end{equation}
where $\vect{V}$ is the non-dimensional velocity field, $P$ is the
non-dimensional pressure, and $Re$ denotes the Reynolds number,
defined with the kinematic viscosity $\nu$ of the fluid, the length
scale $h,$ in our case the channel half-height, and the velocity scale
$U_c$, in our case the centerline velocity of the laminar Poiseuille
profile. The length and velocity scales $h$ and $U_c$ are used to
express any quantity in dimensionless form. The NS equations are then
expressed in a Cartesian coordinate system for the plane channel flow,
with $x, y, z$ (and $U,V,W$) indicating the streamwise, wall-normal
and spanwise directions (and corresponding velocity components). A
sketch of the flow with the employed reference system is shown in
Fig. \ref{fig:channel}.

The boundary conditions for the velocity at the wall are $U=0$ and
$V=0$, but, in contrast to standard Poiseuille flow, the spanwise component is non-zero. Within the general class of streamwise-traveling waves \citep{quadrio-ricco-viotti-2009}, the purely spatially modulated standing wave considered by Viotti \textit{et al.\ }(2009) \cite{viotti-quadrio-luchini-2009} is considered, so that the wall forcing applied at both walls is:
\begin{equation}
W(x) = A \cos \left( \kappa x \right)
\label{eq:forcing}
\end{equation}
where $A$ denotes the maximum forcing amplitude, and $\kappa$ stands
for the forcing wavenumber.

\subsection{The base flow}

It was shown by Viotti \textit{et al.\ }(2009) \cite{viotti-quadrio-luchini-2009} (and later by Quadrio \textit{et al.\ }(2011)
\cite{quadrio-ricco-2011} in the wider context of traveling-wave
forcing) that in the laminar regime the spanwise wall-velocity does
not affect the streamwise flow, which still follows the parabolic
Poiseuille solution, but creates a wall-normal distribution of
spanwise velocity. Hence, the base flow (indicated with an overbar) in
the streamwise direction $\overline{U}(y)=1-y^2$ is identical to the
no-forcing case $A=0$, whereas the spanwise base flow
$\overline{W}(x,y)$, unaffected by the streamwise flow, corresponds to
$Re=0$; it is computed from the spanwise component of the
momentum equation:
\begin{equation}
(1-y^2) \frac{\partial W}{\partial x} = \frac{1}{Re} \Bigl( \frac{\partial^2 W}{\partial x^2 } + \frac{\partial^2 W}{\partial y^2 } \Bigr)
\label{eq:spanflow2}
\end{equation}
with the non-dimensional boundary condition
\begin{equation}
W(x,\pm 1,z,t) = \frac{A}{2} \left( \e^{j\kappa x} + \e^{-j\kappa x} \right) 
\label{eq:spanflow2BC}
\end{equation}
where $j$ is the imaginary unit. Equation \eqref{eq:spanflow2} is linear with streamwise-constant coefficients. By enforcing periodicity and using separation of variables, the $W$ base flow has the form
\begin{equation}
\baseW(x,y) = \Re \left\{ f(y) \e^{j\kappa x} \right\} = \frac{1}{2} \Bigl( f(y) \e^{j \kappa x} + f^*(y) \e^{-j \kappa x} \Bigr) 
\label{eq:Wbaseflow}
\end{equation}
where $\Re$ indicates the real part of a complex quantity, and the
asterisk indicates complex conjugation. The function $f(y)$ is
computed numerically from the following parabolic cylinder equation,
obtained by substituting \eqref{eq:Wbaseflow} into
\eqref{eq:spanflow2} with $f(\pm 1)=A$:

\begin{equation}
f^{''}(y) -\kappa \left[ \kappa + j Re(1-y^2) \right] f(y) =  0.
\label{eq:fsystem}
\end{equation}

Under the hypothesis that the thickness of the transverse Stokes layer
is small compared to $h$, and that streamwise diffusion is negligible
with respect to the wall-normal diffusion, Viotti \textit{et al.\ }(2009) \cite{viotti-quadrio-luchini-2009} have determined the following solution:
\[
f(y) = \mbox{Ai}(-\e^{-j \frac{4}{3} \pi} j y / \delta_x)
\] 
in which $\mbox{Ai}$ represents the Airy function
\citep{abramowitz-stegun-1964}, and $\delta_x= ( \nu / \kappa u_{y,0}
)^{1/3}$ is a representative wall-normal scale of the SSL, defined in
terms of the fluid viscosity $\nu$, the forcing wavenumber $\kappa$,
and the longitudinal wall shear $u_{y,0}$.

\subsection{Linearized equations for the perturbations}

Our interest lies on the dynamics of small perturbations
$u,v,w,p$ to the base flow. After substituting the following decompositions:
\begin{equation}
U = \baseU + u, \qquad V = v, \qquad   W = \baseW + w, \qquad P = \baseP + p,
\end{equation}
into the NS equations \eqref{eq:ns}, and subtracting the base flow equations from the linearized perturbation
equations, obtained by dropping the quadratic terms in the small
perturbations, one obtains
\begin{equation}
\left\{
\begin{array}{ll}
\displaystyle u_x + v_y + w_z = 0, \\
\displaystyle u_t +  \baseU u_x + v \baseU' + \baseW u_z = - p_x + \frac{1}{Re} \nabla^2 u, \\
\displaystyle v_t +  \baseU v_x + \baseW v_z = - p_y + \frac{1}{Re} \nabla^2 v, \\
\displaystyle w_t +  \baseU w_x + u \baseW_x + v \baseW_y + \baseW w_z = - p_z + \frac{1}{Re} \nabla^2 w
\end{array}
\right.
\label{eq:linearized}
\end{equation}
where the prime indicates the wall-normal derivative.

\subsection{The $v-\eta$ formulation}

The usual steps leading to the OSS equations
\citep{schmid-henningson-2001} can be followed here to obtain a
compact expression of the linear dynamics, under the incompressibility
constraint, in terms of two evolution equations for the wall-normal
components $v$ and $\eta = \partial u / \partial z - \partial w /
\partial x$ of the perturbation velocity and vorticity vectors,
respectively. The former equation reads
\begin{dmath}
\frac{\p}{\p t} \nabla^2 v + \left( \baseU \frac{\p}{\p x} + \baseW \frac{\p}{\p z} \right) \nabla^2 v - \left( \baseU'' \frac{\p}{\p x} + \baseW_{yy} \frac{\p}{\p z} \right) v - 2 \baseW_{xy} u_z - 2 \baseW_x \frac{\p}{\p z} \left( u_y - v_x \right) +
\baseW_{xx} v_z = \frac{1}{Re} \nabla^2 \nabla^2 v,
\label{eq:v}
\end{dmath}
while the latter equation can be stated as 
\begin{equation}
\frac{\p}{\p t} \eta + \left( \baseU \frac{\p}{\p x} + 
\baseW \frac{\p}{\p z} \right) \eta +
\left(
   \baseW_x \frac{\p}{\p y} + \baseU' \frac{\p}{\p z} 
 - \baseW_{xy} - \baseW_y \frac{\p}{\p x}
\right) v - u \baseW_{xx} =
\frac{1}{Re} \nabla^2 \eta
\label{eq:eta}
\end{equation}
with boundary conditions of the form 
\begin{equation}
v (\pm 1)=0; \ \ \eta(\pm 1)=0;\ \ \frac{\p v }{\p y} (\pm 1) = 0.
\label{eq:bc_cont}
\end{equation}
As in the standard case, these two equations are complemented by a
differential system which relates $u$ and $w$ to $v$ and $\eta$ via
the continuity equation and the definition of the wall-normal
vorticity. In contrast to the no-forcing OSS case, the two equations
are fully coupled via $\baseW$, and several coefficients are varying
along the streamwise direction.

\subsection{Fourier transform}

The standard OSS system is conveniently Fourier-transformed along the
two homogeneous wall-parallel directions. In our case, this step can
be straightforwardly applied in the spanwise $z$-direction, giving rise to an expansion with spanwise wavenumber $\beta$, since
equations \eqref{eq:v} and \eqref{eq:eta} consist of $z$-independent
coefficients. 
The coefficients' streamwise dependence in equations \eqref{eq:v} and
\eqref{eq:eta} prevents a straightforward Fourier transform in that
direction. However, their functional variation is not generic, as it
derives from the functional form of the base flow \eqref{eq:Wbaseflow}
only; hence the coefficients must be harmonic functions of $x$ with
wavenumber $\kappa$. We follow an approach introduced in Ref.\ \cite{floryan-1997} for the study of the stability of a parallel wall flow modified by periodic blowing and suction, and leverage the sinusoidal variation of the base flow. 
The flow variables are expanded in a
(finite) Fourier series along the streamwise direction as:
\begin{equation}
\check{v}(x,y,t;\beta)    = \sum_{i=-M}^{+M} \hv_i (y,t;\beta)   \e^{j (m+i)\kappa x};
\qquad
\check{\eta}(x,y,t;\beta) = \sum_{i=-M}^{+M} \heta_i (y,t;\beta) \e^{j  (m+i)\kappa x} 
\label{eq:xFT}
\end{equation}
with $M$ as the degree of the spectral expansion of the flow
variables, $i$ as an integer index, and $m \in [0,1)$ as a real number
defining the actual expansion wavenumber $\alpha = (m+i)
\kappa$. The parameter $m$ is referred to as the detuning parameter, since it is used to detune the perturbation against the base flow. In this preliminary work, we set $m=0$. 
Note that Ref.\ \cite{floryan-1997} considered $M=3$, whereas in this work $M$ is much larger and, like the other discretization parameters, is dynamically adapted to each case after a sensitivity analysis. The procedure for setting the discretization parameters is described in Appendix \ref{sec:appendix}.



Equations \eqref{eq:v} and \eqref{eq:eta} can now be
Fourier-transformed along the $x$-direction, introducing the
streamwise wavenumber $\at$ in the process. A generic $x$-dependent
term $q(x)$ is Fourier-transformed as follows:
\[
\hat{q}(\at) = \frac{\kappa}{2 \pi} \int_0^{2 \pi/\kappa} q(x) \e^{-j \at x} dx.
\]

Since in equations \eqref{eq:v} and \eqref{eq:eta} terms with complex
exponentials are of the kind $\e^{j \alpha x}$ and $\e^{\pm j \kappa
  x} \e^{j \alpha x}$, Fourier-transforming in the $x$-direction leads
to integrals of the following general form:
\[
\int_0^{2 \pi/\kappa}  \e^{\pm j \kappa x} \e^{j \alpha x} \e^{-j \at x} dx. 
\]

Owing to the orthogonality of the trigonometric functions, they are proportional to $\delta_{\at,\pm \kappa + \alpha}$, i.e. they are always zero unless $\alpha=\at \mp \kappa$. By introducing the operator $\tilde{\Delta} = \p^2 / \p y^2 - \at^2 - \beta^2$, the equation governing the evolution of $\heta_\at \equiv \heta(y,t; \at,\beta)$ is

\begin{dmath}
\frac{\p }{\p t} \heta_{\at} = - j \at \baseU \heta_{\at} - j \beta
\baseU' \hv_{\at} + \frac{1}{Re} \tilde{\Delta} \heta_{\at} \\
- \frac{j \beta}{2} f \left( 1 - \frac{\kappa^2}{(\at - \kappa)^2 +
 \beta^2} \right) \heta_{\at - \kappa} + \frac{j \beta}{2} f^* \left[ 1
 - \frac{\kappa^2}{(\at + \kappa)^2 + \beta^2} \right] \heta_{\at + \kappa} \\
- \frac{j}{2} \left[ \kappa f \left( 1 + \frac{\kappa (\at -
    \kappa)}{(\at - \kappa)^2 + \beta^2} \right) \frac{\p}{\p y} - f'
  \at \right] \hv_{\at - \kappa} + \frac{j}{2} \left[ \kappa f^*
  \left( 1 - \frac{\kappa (\at + \kappa)}{(\at + \kappa)^2 + \beta^2}
  \right) \frac{\p}{\p y} + f^{*'} \at \right] \hv_{\at + \kappa}
\label{eq:final-eta}
\end{dmath}

Analogously, the equation for $\hv_\at \equiv \hv(y,t; \at,\beta)$ reads
\begin{dmath}
\frac{\p }{\p t} \tilde{\Delta} \hv_\at = - j \at \baseU
\tilde{\Delta} \hv_\at + j \at \baseU'' \hv_\at + \frac{1}{Re}
\tilde{\Delta}\tilde{\Delta} \hv_\at \\
+ \frac{j \beta^2 \kappa}{(\at - \kappa)^2 + \beta^2} \left( f' + f
\frac{\p}{\p y} \right) \heta_{\at - \kappa} - \frac{j \beta^2
  \kappa}{(\at + \kappa)^2 + \beta^2} \left( f^{*'} + f^* \frac{\p}{\p
  y} \right) \heta_{\at + \kappa} \\
- \frac{j \beta}{2} \left[ f \tilde{\Delta}_- - f'' - \kappa f \left(
  2 \at - \kappa \right) + 2 \frac{\kappa (\at - \kappa)}{(\at -
    \kappa)^2 + \beta^2} \left( f' \frac{\p }{\p y} + f \frac{\p^2
  }{\p y^2} \right) \right] \hv_{\at - \kappa} \\
- \frac{j \beta}{2} \left[ f^* \tilde{\Delta}_+ - f^{*''} + \kappa f^*
  \left( 2 \at + \kappa \right) - 2 \frac{\kappa (\at + \kappa)}{(\at
    + \kappa)^2 + \beta^2} \left( f^{*'} \frac{\p }{\p y} + f^*
  \frac{\p^2 }{\p y^2} \right) \right] \hv_{\at + \kappa}
\label{eq:final-v}
\end{dmath}
where $\tilde{\Delta}_\pm = \p^2 / \p y^2 - (\at \pm \kappa)^2 - \beta^2$.

When compared to the standard OSS problem, equations
\eqref{eq:final-eta} and \eqref{eq:final-v} contain several additional
terms related to the spanwise base flow, described by the function
$f(y)$. Terms containing $\hv$ and $\heta$ evaluated at $\at$ are
identical to the OSS equations, whereas the remaining terms represent
effects of disturbances at wavenumbers $\at \pm \kappa$.

\subsection{Numerical discretization}

From here on, a change in notation is introduced. In fact, the
governing equations \eqref{eq:final-eta} and \eqref{eq:final-v} can be
equivalently derived by substituting the modal expansions
\eqref{eq:xFT} into the $v$--$\eta$ equations and by
Fourier-transforming against a testing function of the form
$\e^{j(p+m)\kappa}$, where $p \in [-M,M]$ is an integer such that
$\tilde{\alpha} = (p + m)\kappa$.

Starting from this form, the $y$-dependent terms are discretized via
Chebyshev polynomials on a grid of Gauss-Lobatto nodes
\citep{boyd-1989, canuto-etal-2006, canuto-etal-2007}. The unknown
$\hv_p$ is thus written as:
\begin{equation}
  \hv_p (y) = \sum_{n=0}^{N} {v_{p,n} T_n(y)} 
  \label{eq:chebyshev}
\end{equation}
or, in matrix form, $\hat{\textbf{v}}_p= \textbf{D}_0 \textbf{v}_{p}$,
with $\textbf{D}_0$ as a square $(N+1) \times (N+1)$ matrix. Similar
matrices $\textbf{D}_1$, $\textbf{D}_2$ and $\textbf{D}_4$ represent
the first, second and fourth derivative.

\begin{figure}
\centering
\includegraphics[width=0.6\textwidth]{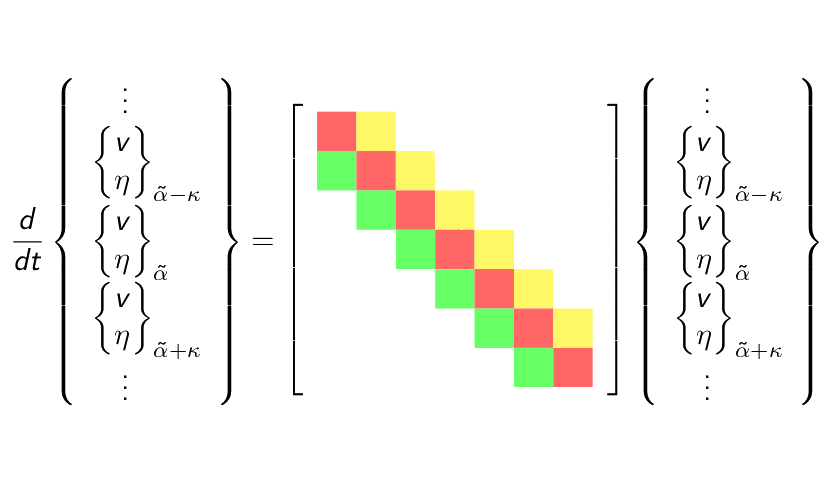}
\caption{Graphical representation of the system $\dot{\textbf{q}} =  \textbf{A}_s \textbf{q}$. The matrix is block-tridiagonal with block size $M \times M$. The red blocks on the diagonal correspond to the classic OSS matrices for each streamwise wavenumber, whereas the adjacent green and yellow blocks describe interactions with contiguous streamwise wavenumbers.}
\label{fig:matrices}
\end{figure}

For a given $p$, equations \eqref{eq:final-eta} and \eqref{eq:final-v}
can be written compactly in the following block form:
\begin{equation}
\label{eq:p_matrixform}
\frac{\partial}{\partial t} \underbrace{\begin{bmatrix} B_{11} & 0 \\ 0 & B_{22} \end{bmatrix}}_{B^{(p)}} \begin{pmatrix} v \\ \eta \end{pmatrix}_p = 
\underbrace{\begin{bmatrix} L_{m_{11}} & L_{m_{12}} \\ L_{m_{21}} & L_{m_{22}} \end{bmatrix}}_{L_-^{(p)}} \begin{pmatrix} v \\ \eta \end{pmatrix}_{p-1} + 
\underbrace{\begin{bmatrix} L_{11} & 0 \\ L_{21} & L_{22} \end{bmatrix}}_{L^{(p)}} \begin{pmatrix} v \\ \eta \end{pmatrix}_p + 
\underbrace{\begin{bmatrix} L_{p_{11}} & L_{p_{12}} \\ L_{p_{21}} & L_{p_{22}} \end{bmatrix}}_{L_+^{(p)}} \begin{pmatrix} v \\ \eta \end{pmatrix}_{p+1}  
\end{equation}
where $\textbf{B}^{(p)}$ is a diagonal matrix, $\textbf{L}^{(p)} $
describes the interaction of $v$ and $\eta$, for a given wavenumber
$\at$, with themselves (red blocks in Fig.\ref{fig:matrices});
$\textbf{L}_-^{(p)}$ and $\textbf{L}_+^{(p)}$ (green and yellow blocks
in Fig.\ref{fig:matrices}) describe the interactions with the previous
and subsequent streamwise wavenumbers, respectively, and contain the
effect of the spanwise base flow.

Equation \eqref{eq:p_matrixform} must be solved for all $p$ of the
truncated modal expansion $[-M, +M]$. Therefore, after grouping the
unknowns $v$ and $\eta$ into a single array of unknowns $\textbf{q}$,
one obtains the following block form:
\begin{equation}
\label{eq:statespaceform}
\frac{\p}{\p t} \Bigl( \textbf{B}_s \textbf{q} \Bigr) = \textbf{L}_s \textbf{q} \hspace{4pc}
\dot{ \textbf{q} } = \textbf{B}_s^{-1} \textbf{L}_s \textbf{q} = \textbf{A}_s \textbf{q}
\end{equation}
where $ \textbf{q}$ is a vector composed of $(N+1)$ pairs of
$v$--$\eta$ components, each at a different $p$ value, so with
dimension $(2M+1) 2 (N + 1)$; $\textbf{B}_s$ and $\textbf{L}_s$ are
block-tridiagonal square matrices, whose dimensions are $(2M + 1) 2(N
+ 1) \times (2M + 1) 2(N + 1)$. The structure of the system matrix
$\textbf{A}_s$ is shown in Fig.\ref{fig:matrices}, which emphasizes its block-tridiagonal structure, with block size $M \times M$.

The boundary conditions \eqref{eq:bc_cont} at the two walls are rewritten as:
\begin{equation}
\hv_p (\pm 1)=0; \ \ \heta_p (\pm 1)=0;\ \ \frac{\p \hv_p }{\p y} (\pm
1) = 0.
\end{equation}
They are readily enforced via appropriate modifications of the
matrices $\textbf{B}^{(p)}$, $\textbf{L}^{(p)} $, $\textbf{L}_-^{(p)}$
and $\textbf{L}_+^{(p)}$. 
Following Ref.\ \cite{schmid-henningson-2001}, the boundary values for $\hv_p$
and $\heta_p$ are set not to exact zero, but to a very small value,
i.e. $10^{-6}$; the extremely fast dynamics associated to this non-zero values does not interfere with the flow dynamics, but the accuracy in computing eigenvalues is improved as the spurious modes associated to the discrete boundary condition are mapped far away in the complex plane.

\subsection{Modal and non-modal stability characteristics}

In what follows, the modal stability of the system will be assessed by
computing the eigenvalues of the system matrix $\textbf{A}_s$ which
describes the linearized dynamics of the system. Due to the size and
sparsity of the matrix, its eigenvalues are most efficiently computed
with the Arnoldi method \citep{trefethen-bau-1997}. The
number of eigenvalues totals $N_{tot} = 2(N + 1) \times (2M + 1)$, but
only a modest fraction of them is of interest here, hence only a
subset $n_{eig} \ll N_{tot}$ is computed. This is achieved by a
truncation operator $\textbf{T}$, a matrix with dimensions $n_{eig}
\times N_{tot}$. The relation $\textbf{q} = \textbf{T} \textbf{x}$
transforms equation \eqref{eq:statespaceform} as
\begin{equation}
  \label{eq:truncation}
  \dot{ \textbf{x} } =\textbf{T}^{-1} \textbf{A}_s \textbf{T}
  \textbf{x}= \boldsymbol{\Lambda} \textbf{x}
\end{equation}
where $\boldsymbol{\Lambda}$ is a diagonal matrix containing the
$n_{eig}$ computed eigenvalues and $\textbf{T}$ contains the
corresponding eigenvectors.

While computing the largest eigenvalue would in principle suffice for
deciding on modal stability or instability, choosing the right size of
the truncation operator is essential to studying the non-modal
stability of the flow (see e.g.\ Refs.\ \cite{reddy-henningson-1993,
  schmid-henningson-2001, criminale-jackson-2003}). The kinetic energy
density $e$ of an infinitesimal perturbation for a given $p$ is
written as
\begin{equation}
  e(p,\beta,t) = \frac{1}{2} \left( \frac{1}{2k^2} \int_{-1}^{1}
  \hat{\textbf{q}}_p^H
\begin{bmatrix} k^2+D_1^2 & 0 \\ 0 & 1 \end{bmatrix}
\hat{\textbf{q}}_p dy \right)
\end{equation}
with $ k^2 = \at^2 + \beta^2$; the superscript $H$ stands for
Hermitian, i.e., the complex conjugate transpose.

After a Chebyshev expansion of $\textbf{q}_p$ and after computing the
integral weights using the Clenshaw-Curtis quadrature formula
\citep{clenshaw-curtis-1960, imhof-1963, hanifi-schmid-hennigson-1996,
  maleknejad-lotfi-2005}, one obtains
\begin{equation}
  e(p,\beta,t) = \begin{pmatrix} \hat{\textbf{v}}_p
    \\ \hat{\boldsymbol{\eta}}_p \end{pmatrix}^H \underbrace{\Biggl(
    \frac{1}{4 k^2}
    \begin{bmatrix} k^2 \textbf{D}_0^H \textbf{W} \textbf{D}_0 + \textbf{D}_1 \textbf{W} \textbf{D}_1 & \textbf{0} \\ \textbf{0} & \textbf{D}_0^H \textbf{W} \textbf{D}_0 \end{bmatrix}}_{\textbf{Q}^{(p)}} \Biggr)
\begin{pmatrix} \hat{\textbf{v}}_p  \\ \hat{\boldsymbol{\eta}}_p  \end{pmatrix}
\end{equation}
where $\textbf{W}$ is a diagonal matrix containing the integral
weights, and $\textbf{Q}^{(p)}$ represents the energy weight matrix.

Once the full modal expansion $p\in [-M,M]$ is considered, the square
of the energy norm of $\textbf{q}$ becomes:
\[
{\| \textbf{q}(t) \|}_E^2  = \textbf{q}^H \textbf{Q}_s \textbf{q}  
\]
where $\textbf{Q}_s$ is the block-diagonal positive definite energy
weight matrix for all indices $p$ in the modal expansion. The energy
norm can be rewritten using the truncation operator $\textbf{T}$:
\[
{\| \textbf{q}(t) \|}_E^2 = \textbf{x}^H \textbf{T}^H \textbf{Q}_s
\textbf{T} \textbf{x} = \textbf{x}^H \bar{\textbf{Q}} \textbf{x}
\]
where $\bar{\textbf{Q}} = \textbf{T}^H \textbf{Q}_s \textbf{T}$ can be
further decomposed via a Cholesky decomposition $\bar{\textbf{Q}} =
\bar{\textbf{C}}^H \bar{\textbf{C}}$ to result in
\[
{\| \textbf{q}(t) \|}_E^2 = {\| \bar{\textbf{C}} \textbf{x}(t)
  \|}_2^2.
\]

The transient energy growth eventually becomes
\begin{equation}
  G(t) = \max_{ \textbf{q}_0 \not= 0 } \frac{{\| \textbf{q}(t) \|}_E^2
  }{{\| \textbf{q}_0 \|}_E^2} =
  \max_{ \textbf{x}_0 \not= 0 } \frac{{\| \bar{\textbf{C}} \e^{\Lambda
        t}\textbf{x}_0 \|}_2^2 }{{\| \bar{\textbf{C}} \textbf{x}_0
      \|}_2^2} = { \| \bar{\textbf{C}} \e^{\Lambda t} \bar{\textbf{C}}
    ^{-1} \| }_2^2
\label{eq:gFunc}
\end{equation}

Lastly, the spatial shape of the optimal perturbation
$\hat{\textbf{q}}_{in}$, and the corresponding spatial shape at the
time of the largest energy growth $\hat{\textbf{q}}_{out}$, are given
by:
\begin{align}
  \label{eq:optimal}
  \hat{\textbf{q}}_{in} = \textbf{D}_0 \textbf{T} \bar{\textbf{C}}
  ^{-1} \textbf{v}_1 , \qquad \hat{\textbf{q}}_{out} = \textbf{D}_0
  \textbf{T} \bar{\textbf{C}} ^{-1} \textbf{u}_1 ,
\end{align}
where $ \textbf{v}_1 $ and $ \textbf{u}_1 $ are the first right and
left singular vectors of the matrix $ \bar{\textbf{C}} \e^{\Lambda t}
\bar{\textbf{C}} ^{-1}$, respectively.

\subsection{Validation}
\label{sec:val}

The MATLAB numerical toolkit developed for the stability analysis is
first validated against the results of a non-linear DNS solver. The
code, introduced in Ref.\ \cite{luchini-quadrio-2006}, where full details are available, solves the
incompressible Navier--Stokes equations with mixed spatial
discretization, with Fourier expansions in the homogeneous directions
(where the pseudo-spectral approach is used) and compact, fourth-order
explicit finite-difference schemes in the wall-normal direction.

\begin{figure}
\centering
\includegraphics[trim=3cm 2cm 3cm
  3cm,clip,width=0.8\textwidth]{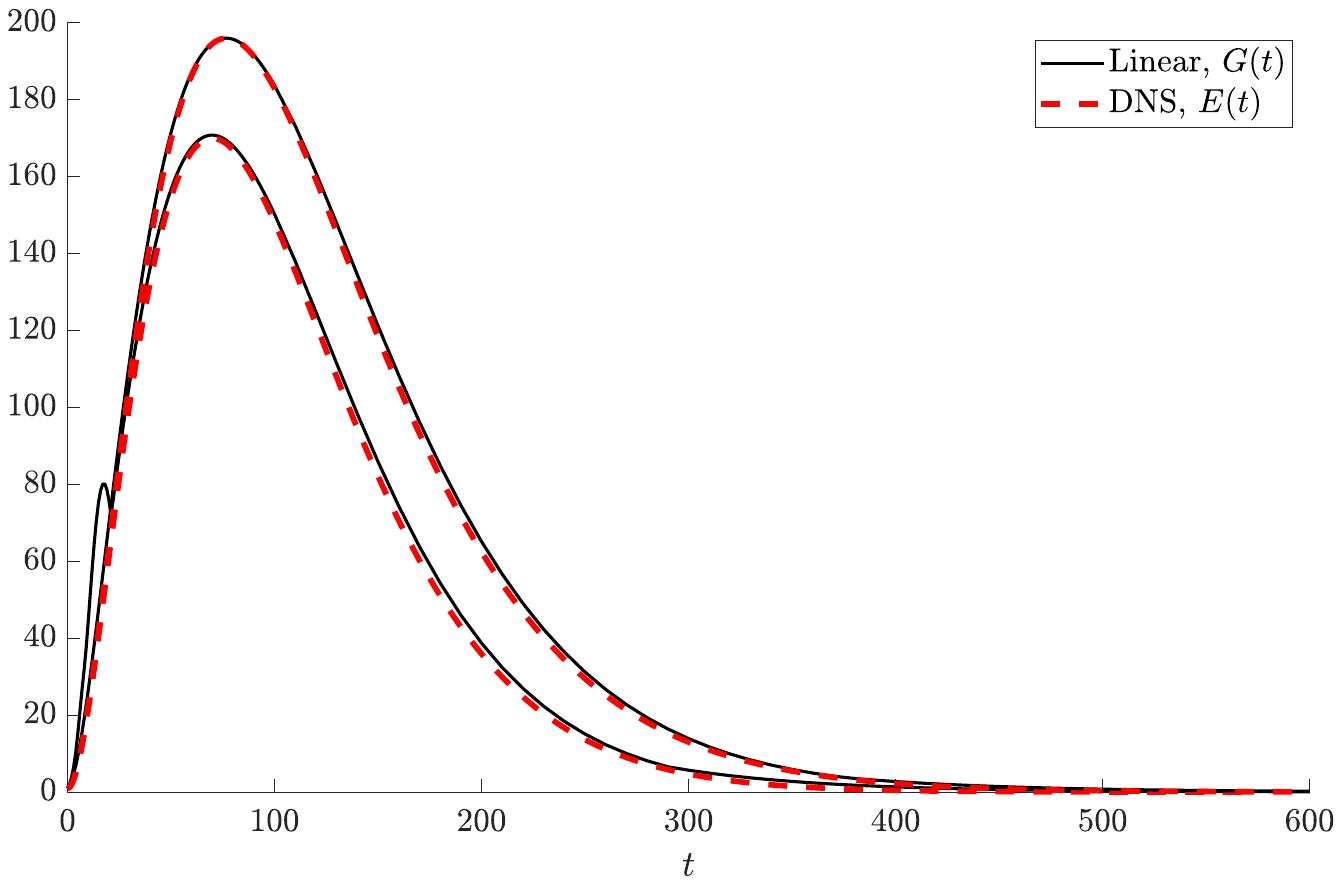}
\caption{Comparison between the transient energy growth function
  $G(t)$ computed by the linear code (black continuous line) and the DNS-computed temporal
  evolution of the perturbation energy $E(t)$ (red dashed line). Reference conditions
  are $Re=1000$, $\kappa=1$ and $\beta=2$. The higher pair of curves
  corresponds to $A=0$ (no control), and the lower one to $A=0.1$.}
\label{fig:nonmodal-validation}
\end{figure}

The transient energy growth function $G(t)$ computed by the stability
code is compared with the DNS-computed temporal evolution of the
energy $E(t)$ of the optimal initial condition. The test is carried
out for $Re=1000$, $\kappa=1$, $\beta=2$. According to the
discretization criteria described later in
Sec.\ref{sec:discretization} and Appendix \ref{sec:appendix}, 
the discretization parameters are set to
$N=80$, $M=10$ and $A=0$ or $A=0.1$. The DNS is carried out with 101
points in the wall-normal direction, $32$ Fourier modes in the
streamwise direction, and 16 modes in the spanwise direction; the size of the
temporal step is computed via the Courant-Friedrich-Lewy (CFL)
condition, by enforcing $CFL_{max}=0.1$; it is verified that the
outcome of the DNS is insensitive to further refinements in the
temporal and/or spatial discretizations. The initial field for the DNS
is made by the base flow, to which the optimal initial condition
obtained by the non-modal stability code is added, after rescaling the
amplitude to remain within the limits of linearity. Figure
\ref{fig:nonmodal-validation} compares the temporal history $E(t)$ of
the DNS-computed energy with the transient growth function $G(t)$ for the uncontrolled and controlled cases. 
It can be observed that, as required by their definition, $G(t) \ge E(t)$ at all times; the two functions coincide at time $t=t_{Gmax}$. 
The controlled case with this particular choice of parameters exhibits a reduced maximum transient growth.

%% file: discretization.tex
\section{Parameters and computational procedures}
\label{sec:discretization}

The present multi-parameter study considers the effects of the
following nine parameters, either physical or related to the
discretization of the problem:

\begin{enumerate}
\item $Re$, the Reynolds number of the flow; 
\item $A$, the maximum amplitude of the wall forcing in equation
  \eqref{eq:forcing};
\item $\kappa$, the streamwise wavenumber of the wall forcing in
  equation \eqref{eq:forcing};
\item $\beta$, the spanwise wavenumber of the perturbation;
\item $N$, the number of Chebyshev polynomials in equation \eqref{eq:chebyshev}
\item $M$, the truncation factor in the modal expansion \eqref{eq:xFT}
\item $n_{eig}$, the number of eigenvalues retained in the Arnoldi algorithm after truncation;
\item $\Delta t$, the temporal step for the discrete evaluation of $G(t)$;
\item $T_{end}$, the time at which the computation of $G(t)$ is stopped.
\end{enumerate}

The first four parameters listed above are of a physical nature, and
their range defines the breadth of the study. The remaining parameters
are discretization parameters which impact the reliability of the
results and the computational cost of the study. 
The vast range of explored physical parameters, together with the anticipated
variable behavior of the system within it, requires efficiency and
sensitivity considerations with regard to the chosen discretization.


Description of the choice of the discretization parameters is deferred to Appendix \ref{sec:appendix}.
As far as the physical parameters are concerned, only three
subcritical values of the Reynolds number are considered in this
study, namely $Re=500$, $Re=1000$ and $Re=2000$. The wall forcing is
defined by two parameters: its dimensionless amplitude $A$ is varied
between 0 and 1 in increments of 0.1 (resulting in 11 values), and its
dimensionless wavenumber $\kappa$ is varied between 0.5 and 5 in
increments of 0.25 (yielding 19 values). The spanwise wavenumber
$\beta$ is varied between 0 and 5, with 18 non-equispaced values
specifically selected to focus on the most interesting regions. 
Note that $\beta=0$ corresponds to the most unstable two-dimensional waves predicted by the modal theory. $A=1$ implies a maximum spanwise velocity equal to the centerline Poiseuille velocity, and $\kappa=1$ implies a forcing wavelength that is $\pi$ times the channel heigth.
Overall, 11,286 cases are computed.  The total computational cost is thus
considerable, and exceeds 10000 core hours; a workstation equipped with an Intel i7
CPU with 6 cores has been used.

%% file: results.tex
\section{Results}
\label{sec:results}

\subsection{Modal stability}
\label{sec:results-modal}

The modal stability  characteristics of the flow are evaluated in terms of the real part of the least stable eigenvalue $\lambda_1$ of the system matrix $\textbf{A}_s$. The effect of wall forcing, represented through the physical parameters $Re$, $\kappa$, $A$ and $\beta$, on the modal stability properties is quantified via the attenuation-rate increase, expressed via the ratio $R_{mod}$, defined as:
\[
R_{mod} = \frac{\Re(\lambda_1)}{\Re(\lambda^{ref}_1)},
\]
where $\lambda^{ref}_1$ is the least-stable eigenvalue of the unforced
Poiseuille flow. Since the considered $Re$-numbers are subcritical,
$\Re(\lambda_1^{ref})<0$, and a positive effect of the forcing in the
direction of increased stability margin implies $R_{mod}>1$, in analogy with the drag reduction rate used in the turbulent case to assess the effectiveness of the forcing. The
discretization parameters are set according to the criteria described
in Appendix \ref{sec:appendix}.



\begin{table}
\centering
\begin{tabular}{ccccc} 
$Re$ & $\kappa$ & $\beta$ & $A$ & $R_{mod}$ \\ 
& & & & \\
500  & 2.25 &  0.7  & 1  &   1.9380\\
1000 & 2.00 &  0.5  & 1  &    2.1725\\
2000 & 3.00 &  0.5  & 1  &    2.3636\\ 
\end{tabular}
\caption{Largest relative increase $R_{mod}$ of the modal stability
  margin, and the corresponding forcing parameters at which it is
  observed.}
\label{tab:modal}
\end{table}

\begin{figure}
  \centering
  \includegraphics[trim=2.5cm 9cm 3cm
    9.5cm,clip,width=0.6\textwidth]{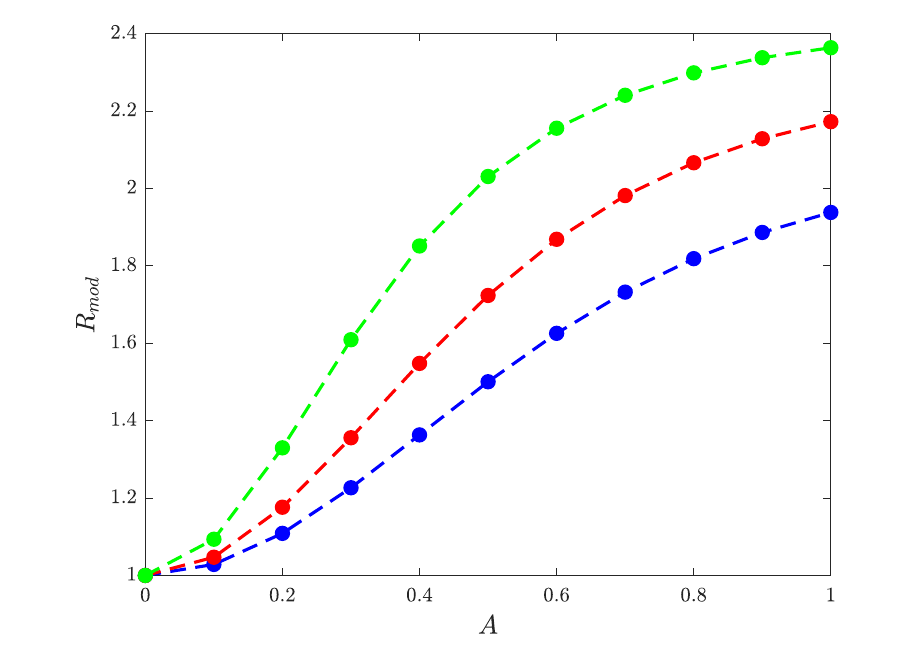}
  \caption{Dependence of $R_{mod}$ on the forcing amplitude $A$ at
    optimal $\kappa$ and $\beta$ reported in table \ref{tab:modal} for
    $Re=500$ (blue), $Re=1000$ (red), and $Re=2000$ (green).}
  \label{fig:modal}
\end{figure}

For the various values of $Re$, table \ref{tab:modal} reports the
values of $A$, $\beta$ and $\kappa$ which have been found to yield the
largest increase of $R_{mod}$. The results are $Re$-dependent, with
$R_{mod}$ going from $1.9380$ at $Re=500$ ($\lambda_1$ decreases from -0.00591 to -0.01146) to $2.3636$ at
$Re=2000$ ($\lambda_1$ decreases from -0.00135 to -0.00321). 
The stabilizing effect of spanwise forcing is significant,
since at $Re=2000$ the negative real part of the least-stable
eigenvalue increases (in absolute value) by more than 2.3 times. It
should come as no surprise that the effectiveness of the forcing
depends on its amplitude $A$: indeed, in table \ref{tab:modal} the
largest $R_{mod}$ are consistently obtained for the largest forcing
amplitude tested, i.e. $A=1$. Once the optimal values of $\kappa$ and
$\beta$ are determined, figure \ref{fig:modal} depicts how $R_{mod}$
changes as a function of $A$. For all Reynolds number studied,
$R_{mod}$ is quite similar to the turbulent drag reduction rate, and exhibits a monotonic growth from the uncontrolled
case with $R_{mod}=1$, then saturating at large amplitudes. 
Especially at higher $Re$, the effect of
forcing is already noticeable at rather small forcing intensity: for
example, at $Re=2000$ a forcing with $A=0.5$ provides more than 85\%
of the benefit achievable at $A=1$. This observation is important in
view of the rapid increase of the energetic cost of the forcing.

\begin{figure}
  \centering
  \includegraphics[trim=1cm 0.88cm 1cm 1cm,clip,width=0.9\textwidth]{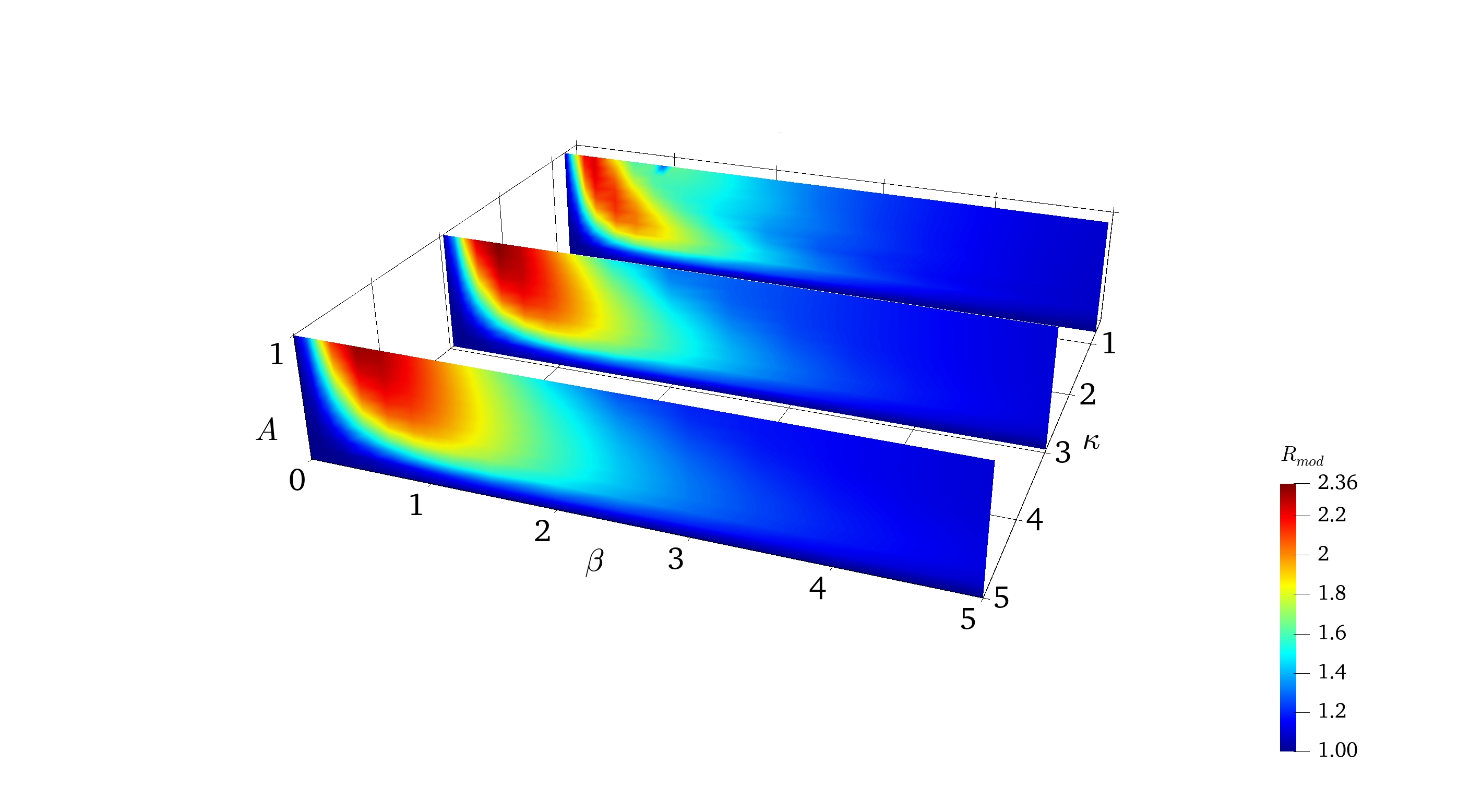}
  \caption{Variation of $R_{mod}$ at $Re=2000$ in three selected
    planes at $\kappa=0.75, 2.75, 5$. }
\label{fig:modal_k}
\end{figure}

To examine how the results depend on $\kappa$, we first observe that,
in table \ref{tab:modal}, the optimal $\kappa$ does not vary much with $Re$. 
Indeed, the optimum with respect to $\kappa$ is
rather flat. As an example, at $Re=2000$ figure \ref{fig:modal_k}
plots $R_{mod}$ in $A$-$\beta$-planes taken at three selected values
of $\kappa$, namely $\kappa=0.75$, $\kappa=2.75,$ and $\kappa=5$. The
plot demonstrates the weak dependence of $R_{mod}$ on the forcing
wavenumber, especially in the vicinity of the optimum value
$\kappa=3$.

\subsection{Non-modal stability}

\begin{table}
  \centering
  \begin{tabular}{ccccc} 
    $Re$ & $\kappa$ & $\beta$ & $A$ & $R_{nmod}$ \\ 
    & & & & \\
    500  & 1.25 &  1.5  & 1  &    2.8877\\
    1000 & 0.75 &  2.5  & 1  &    3.3670\\
    2000 & 0.75 &  1.5  & 1  &    3.6140\\ 
  \end{tabular}
  \caption{Largest decrease of the transient growth $R_{nmod}$ and
    corresponding forcing parameters.}
  \label{tab:nonmodal}
\end{table}

To shed light on the forcing-induced modifications to the short-term
stability properties of the flow, we now proceed to consider the transient growth
function $G(t)$ and its maximum $G_{max}$. $G_{max}$ is the maximum
possible relative amplification of the initial perturbation energy,
occurring at time $t_{max}$ for a specific initial condition referred
to as the optimal input. We quantify the effect of the forcing on the non-modal stability by computing the ratio $R_{nmod}$ defined as:
\[
R_{nmod} = \frac{G_{max}^{ref}}{G_{max}} ,
\]
where $G_{max}^{ref}$ is the maximum transient growth for the
reference Poiseuille flow. As in the modal analysis, the
discretization parameters are set according to the criteria discussed
in Appendix \ref{sec:appendix}. Table \ref{tab:nonmodal} presents the
optimal parameters $A$, $\kappa$ and $\beta$ which have been found
 to provide the largest $R_{nmod}$. As for modal
stability, changes in $R_{nmod}$ are Reynolds-number-dependent, and
the maximum reduction of the transient energy growth ranges from $65\%$
at $Re=500$ (with $G_{max}$ decreasing from 43.39 to 15.03) to $72\%$ at $Re=2000$
(with $G_{max}$ decreasing from 689.52 to 190.80).

\begin{figure}
  \centering
  \includegraphics[trim=2.5cm 9cm 3cm
    9.5cm,clip,width=0.4\textwidth]{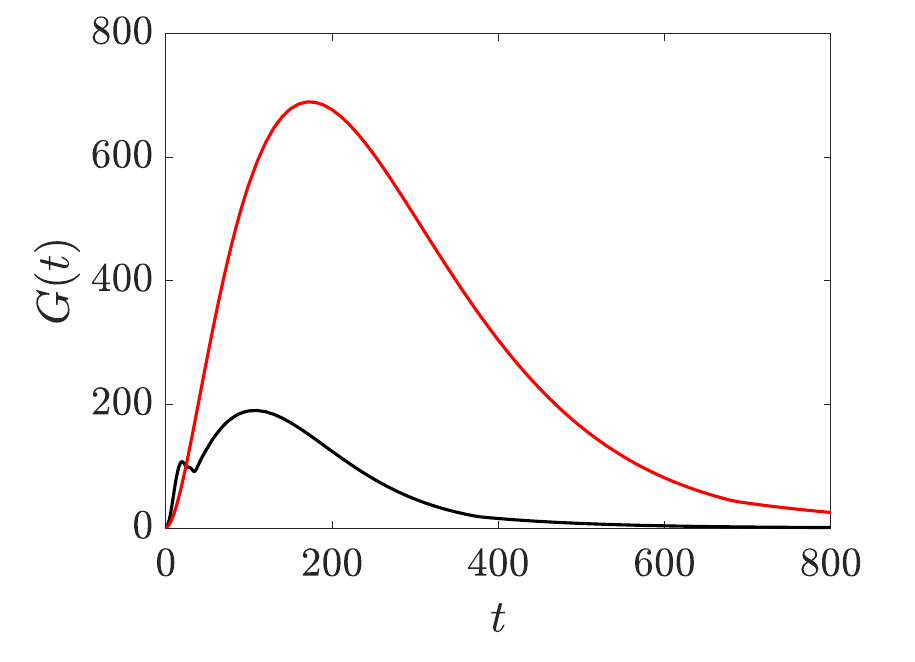}
  \includegraphics[trim=2.5cm 9cm 3cm
    9.5cm,clip,width=0.4\textwidth]{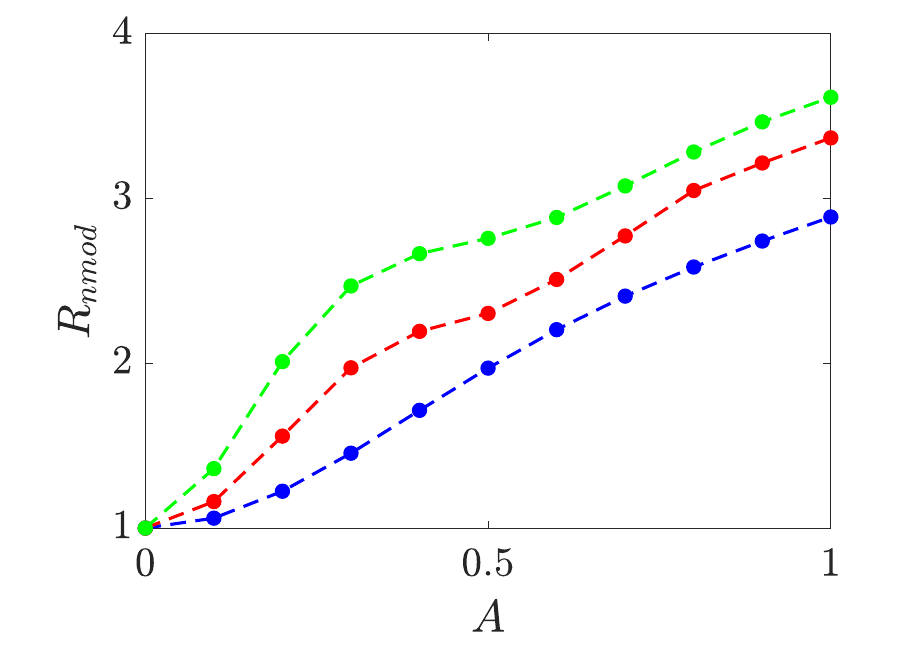}
  \caption{Left: transient energy growth $G(t)$ for $Re=2000$ (black) for
    the optimal values of $\beta=1.5$ and $\kappa=0.75$, compared to
    the unforced case (red). Right: dependence of $R_{nmod}$ on the forcing amplitude $A$, for $Re=500$ (blue), $Re=1000$ (red), and $Re=2000$ (green). }
\label{fig:Gfunc}
\end{figure}

At the optimal values $(\kappa, \beta, A)$ reported in table
\ref{tab:nonmodal}, the transient energy growth with and without
forcing is compared in figure \ref{fig:Gfunc} (left). The transient growth is
clearly inhibited by the spanwise forcing, and indeed the maximum
energy amplification decreases from about 700 times the initial energy
to less than 200 times; moreover, the maximum amplification occurs at
an earlier time. At least for this specific case, $G(t)$ for the
forced case is above the reference curve for very short times.

The role of the two parameters $A$ and $\kappa,$ defining the wall
forcing, is similar to the one discussed previously for modal
stability. The forcing amplitude directly affects the amount of
transient growth reduction, as clearly shown in figure \ref{fig:Gfunc} (right),
while the forcing wavenumber $\kappa$ has a lesser effect: as the wavenumber exceeds the
optimal values identified in table \ref{tab:nonmodal} (which are close
to the minimum value of $\kappa=0.5$ considered in the present study),
the effectiveness of the forcing does not exhibit a significant or
rapid decrease.

\begin{figure}
  \centering
  \includegraphics[trim=2cm 3cm 1cm
    1.5cm,clip,width=1\textwidth]{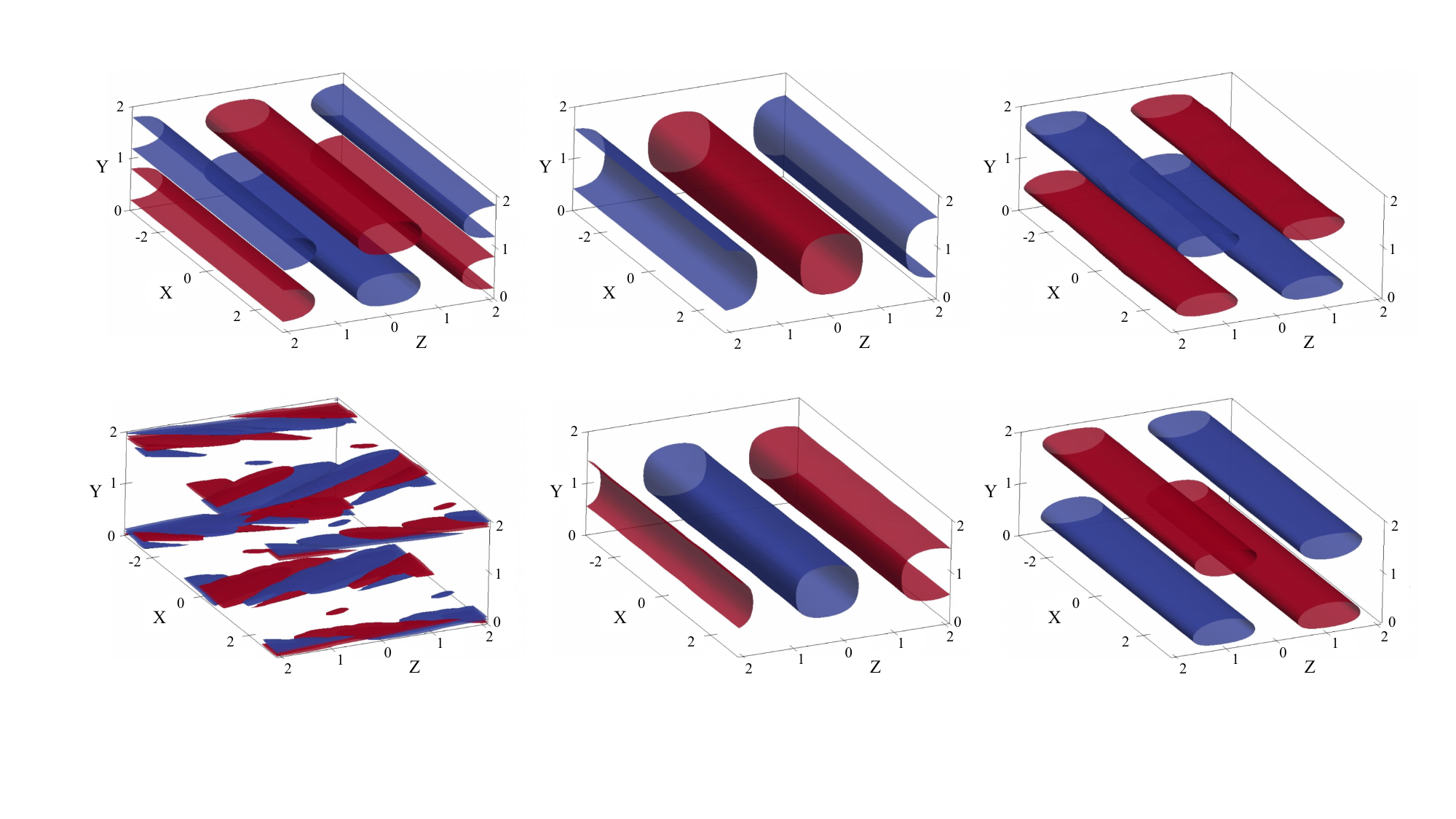}
  \caption{Optimal initial condition computed for $Re=2000$,
    $\kappa=1$ and $\beta=1.5$: unforced flow at $A=0$ (top) and
    forced flow at $A=1$ (bottom). For each velocity component ($u, v,
    w$ from left to right) isosurfaces at $\pm 60\%$ of their
    respective maxima are visualized. After normalizing the
    perturbation to unit maximum amplitude, the $u,v,w$ components
    of the maximum are $0.11, 2.68, 3.79$ for $A=0$, and $0.74, 2.63,
    3.79$ for the $A=1$ case.}
\label{fig:optinp}
\end{figure}

\begin{figure}
  \centering
  \includegraphics[trim=2cm 3cm 1cm
    1.5cm,clip,width=1\textwidth]{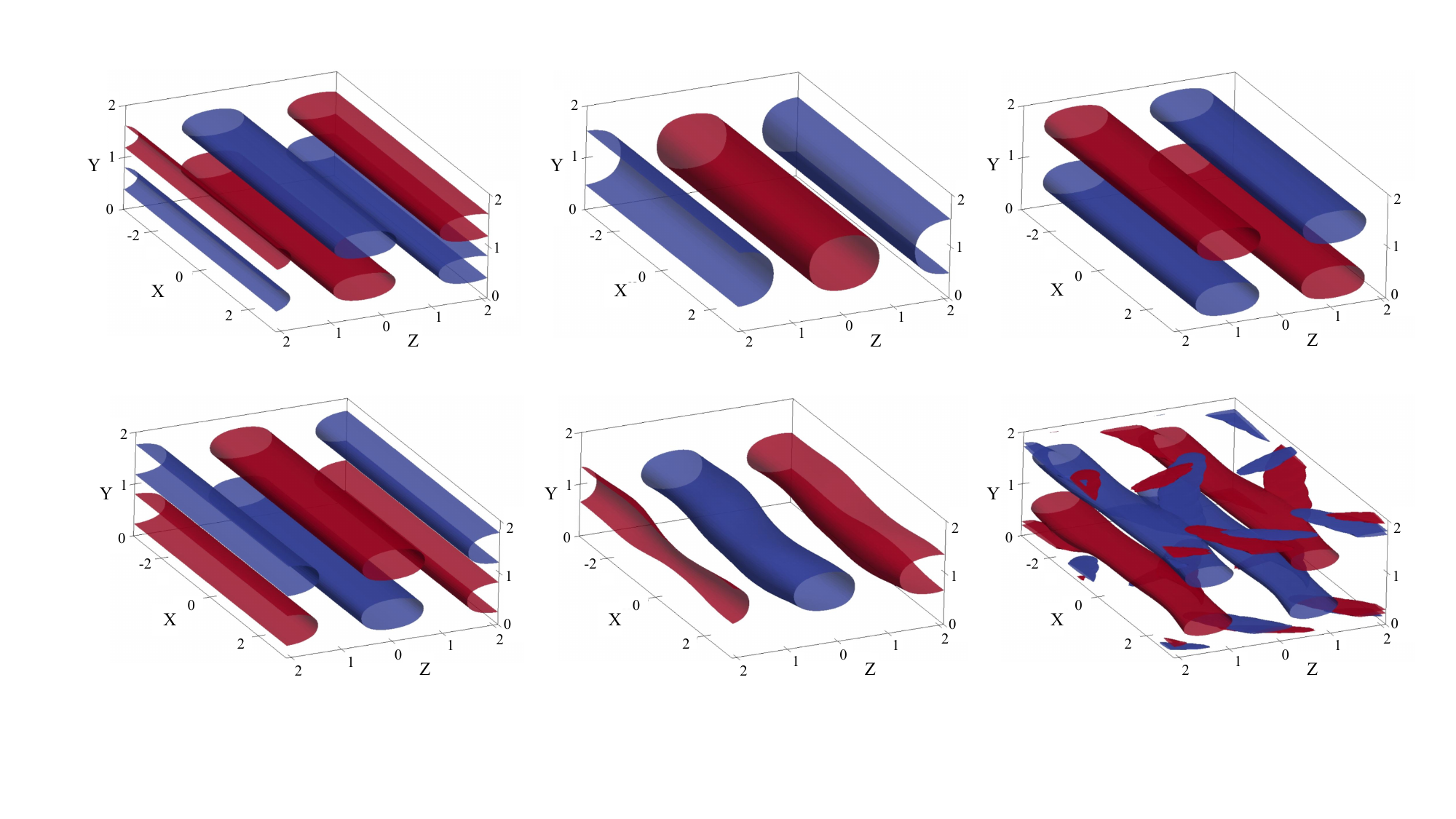}
  \caption{Evolution of the optimal initial condition at the time of
    maximum energy growth, computed for $Re=2000$, $\kappa=1$ and
    $\beta=1.5$: unforced flow at $A=0$ (top) and forced flow at $A=1$
    (bottom). For each velocity component ($u, v, w$ from left to
    right) isosurfaces at $\pm 60\%$ of their respective maxima are
    visualized. The $u,v,w$-components change by a factor of
    $481,0.47,0.40$ for $A=0$, and by a factor of $44,0.51,0.46$ for
    $A=1$ with respect to the initial condition.}
  \label{fig:optout}
\end{figure}

Nonmodal stability theory readily provides access to the shape of the
perturbation that triggers the maximum transient energy growth. The
optimal input (and the corresponding optimal output), discussed in section \ref{sec:math}
 and obtained by means of a singular values
decomposition, are examined in physical space in figures
\ref{fig:optinp} and \ref{fig:optout}.

In particular, figure \ref{fig:optinp} describes how the optimal
initial condition is affected by the forcing. The top row refers to
the reference case with $A=0$ and confirms that, for plane Poiseuille
flow, the optimal initial condition is a streamwise-constant pair of
large-scale space-filling rolls. Indeed, in Fourier space such a
perturbation is fully represented by the mode $p=0$, leading to an
effective streamwise wavenumber $\alpha=0$. When the spanwise forcing
is active with $A=1$ (bottom row), the optimal initial condition is
characterized by different streamwise wavenumbers. In particular, the
morphology of the streamwise component of the optimal initial
condition reveals the fundamental mode of the forcing, with superimposed
higher harmonics. A striking difference to the unforced case is that
the $u$-component, although not entirely two-dimensional, is predominantly structured along the spanwise direction, i.e. orthogonal to the unforced case. Another
significant difference is the fact that the perturbation becomes
highly localized near the wall, and remains confined within
the high-shear region produced by the spanwise Stokes layer described by the base flow. 
Though barely visible at this $Re$, a streamwise modulation can be observed in the
spanwise component as well, whose overall shape however still
resembles very much that of the unforced flow. The vertical component appears to
be essentially unaffected by the forcing. In terms of relative intensity, the three
velocity components are comparable with/without control; while the amplitudes of the wall-normal and spanwise components are unchanged, that of the longitudinal component, which is the smallest, grows by about six times in the controlled case.

If attention is drawn to the temporal evolution of the optimal initial
condition, shown in figure \ref{fig:optout} at the time of maximum
energy amplification, similar effects are evident regarding its spatial shape;
the spanwise forcing at the wall introduces a streamwise modulation
with higher harmonic components. The modulation is however more
reduced in comparison to the initial condition, and -- at least for
the case under consideration -- it appears that at $t=t_{max}$ the
perturbation has developed the same qualitative shape in both
cases. The reorientation of the $u$-component from the spanwise to the
streamwise direction is attributed to an Orr-type mechanism
\citep{orr-1907}: the initial disturbance is characterized by a flow
pattern opposed to the mean shear which, as time evolves, is tilted
into the shear direction.

To explain how the transient growth is altered, one has to consider
the energy levels of the perturbation. Within the present linear
setting the absolute value of the perturbation amplitude is
inconsequential, and perturbations are normalized to unit maximum
absolute value. However, the relative values of the various components
and their growth rates can be informative. In the reference flow, the
three velocity components of the initial condition at $Re=2000$ are
relatively balanced, but the transient growth affects them very
differently. In fact, the relative increment of the amplitude of the
streamwise component is 481 times the initial value (this figure is
obviously $Re$-number-dependent), whereas the $v$ and $w$ components
decrease by factors of 0.47 and 0.46, respectively. With spanwise
forcing, however, the picture is drastically changed. The evolution of
the $v$- and $w$-components is similar to the unforced case, as they
reduce to 0.51 and 0.46 of their initial value for $v$ and $w,$
respectively. However, the growth of the streamwise component is
severely inhibited, and at $t=t_{max}$ the increase of the streamwise
velocity component is only 44 times, instead of 481 times.

\subsection{Discussion}

\begin{figure}
\centering
\includegraphics[trim=3cm 9.2cm 3.5cm 9cm,clip,width=0.45\textwidth]{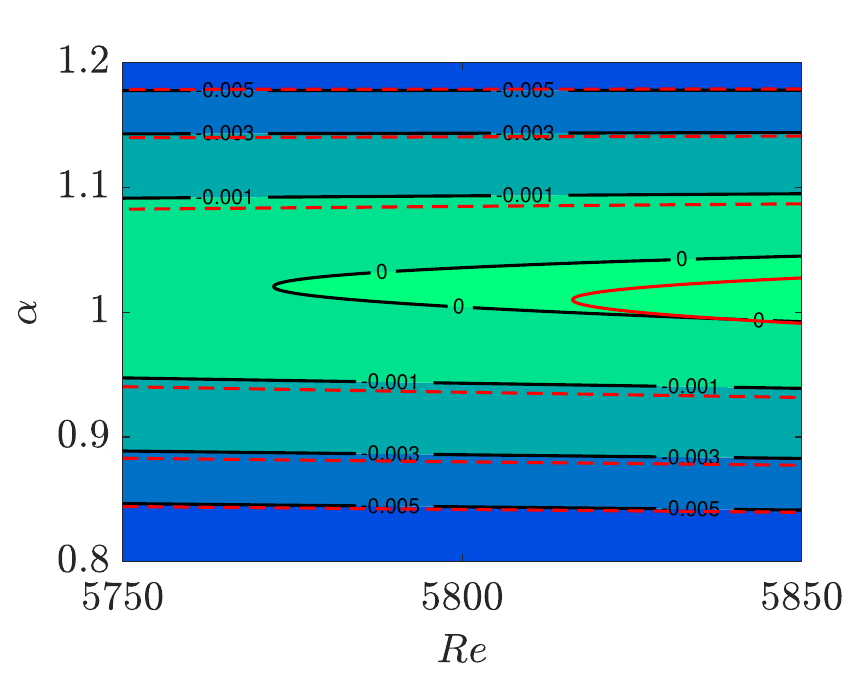}
\includegraphics[trim=2.5cm 9cm 3cm 10cm,clip,width=0.45\textwidth]{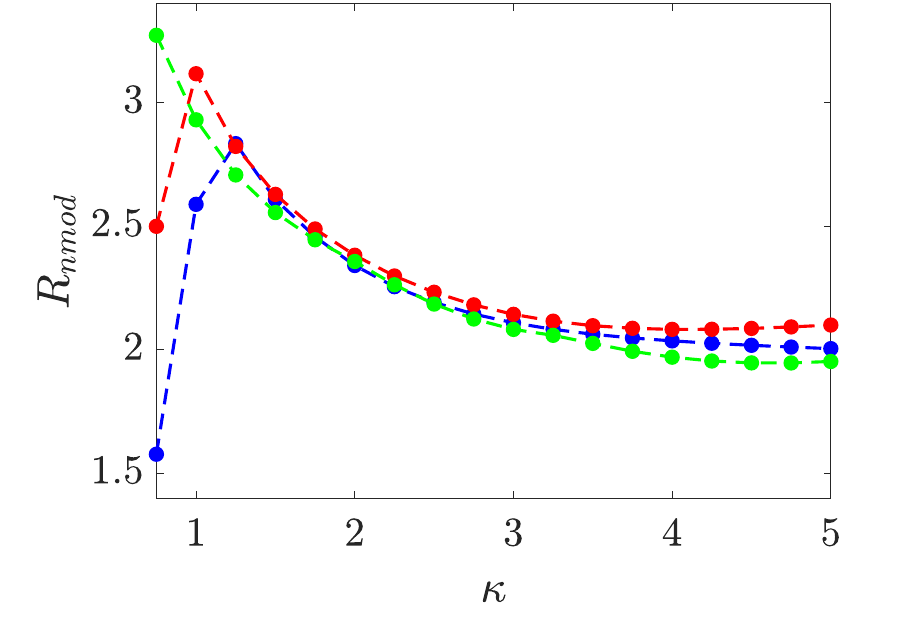}
\caption{Modal and non-modal stability characteristics computed by taking $\beta$ as a free parameter to select the worst case. Left: neutral stability curves for the unforced flow (contour plot and black lines) and the forced flow (red lines) for $A=1$ and $\kappa=3$. Right: dependence of $R_{nmod}$ on $\kappa$ at $A=1$ and $Re=500$ (blue), $Re=1000$ (red) and $Re=2000$ (green).}
\label{fig:betaworstcase}
\end{figure}

The largest $R_{mod}$ and $R_{nmod}$
identified in this study correspond to the best performance observed
in the parameter space ($A$, $\kappa$, $\beta$) for a given $Re$. However, it
is important to realize that the amplitude and
wavenumber are indeed specifiable parameters of the forcing, but the
spanwise wavenumber $\beta$ of the perturbation, describing a generic small
disturbance, is not a control
parameter. Furthermore, background
disturbances are realistically assumed to contain all frequencies. Hence, to
appreciate the true potential of spanwise forcing, one should compare
the neutral curve of the unmanipulated Poiseuille flow with a neutral curve for the
controlled Poiseuille flow, where all values of $\beta$ are
considered and the worst-case scenario is selected. This is shown in
figure \ref{fig:betaworstcase} (left), where a forcing defined by $\kappa=3$
and $A=1$ is selected from the optimal parameters identified above. For
the neutral curve we scan through $Re$ from 5750 to 5850 by increments
of 5, and through the wavenumber $\alpha$ from 0.8 to 1.2 by
increments of 0.05. Every point in the $Re$-$\alpha$ plane for the
forced flow results from a scan over all available
spanwise wavenumbers $\beta$. For the unforced flow the critical value $Re_c$ of the Reynolds number is the well-known $Re_c=5772$; for $A=1$, this threshold is increased, but only marginally so, to $Re_c=5816$. The increment is not particularly relevant, in contrast to the optimal situation discussed above in \ref{sec:results-modal}. In line with the minimal change of $Re_c$, the shape of the most unstable eigenmode, i.e. the linearly unstable Tollmien--Schlichting wave, is also not altered significantly by the forcing. This means that the spanwise forcing is indeed capable of hindering the growth of certain spanwise perturbations, but not of all of them. Once the global effect of the forcing is considered in a scenario where perturbations of any wavenumber $\beta$ are present, the improvement is significantly smaller. 

However, in the context of highly subcritical flows the significance of this result is limited, since the non-modal stability characteristics are way more important. With non-modal stability, the overall picture improves substantially. The
maximum transient growth is significantly inhibited by spanwise
forcing, regardless of the considered perturbation. As shown in table
\ref{tab:nonmodal}, the ratio $R_{nmod}=G_{max}^{ref}/G_{max}$
increases from $2.89$ at $Re=500$ to $3.61$ at $Re=2000$. Following the
discussion above, it is interesting to assess how the forcing improves
the stability characteristics by finding the value of $\beta$
which yields the minimum $R_{nmod}$. Figure \ref{fig:betaworstcase} (right) plots
how $R_{nmod}$ depends on $\kappa$, at $A=1$ and for the three
Reynolds numbers, after selecting the $\beta$ that yields the worst performance. 
Unlike the modal case, the beneficial effects of the
spanwise forcing on the nonmodal stability characteristics are
significant. 
Figure \ref{fig:betaworstcase} (right) confirms again that, once $\kappa$ remains at or
above the optimal values, the forcing remains rather effective, and
consistently guarantees at least halving the transient energy
growth, regardless of $Re$ and the wavenumber of the spanwise
perturbation. The case $\beta=0$ is the worst scenario only
in the uncontrolled case; for each data point presented in Figure \ref{fig:betaworstcase}, a comprehensive parametric study is conducted to determine the value of $\beta$ that yields
the worst modal/non-modal improvement. Overall, the spanwise forcing appears to have little or no effect on the linearly unstable Tollmien--Schlichting waves detected by modal stability, but interacts favourably with the lift-up effect.

\begin{figure}
\centering
\includegraphics[trim=2.5cm 9cm 3cm 9.5cm,clip,width=0.6\textwidth]{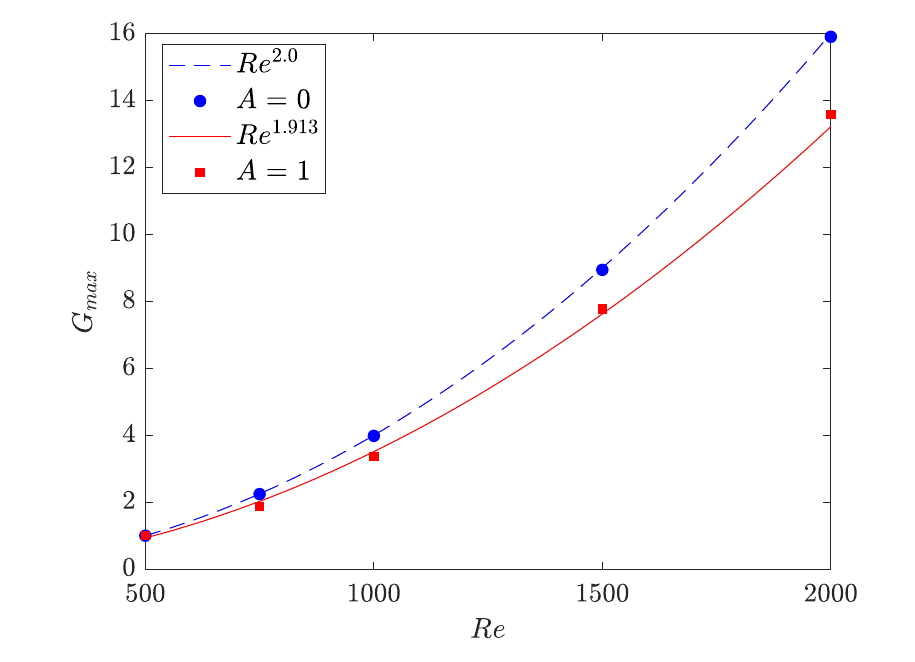}
\caption{Dependence of the maximum transient growth $G_{max}$ on the Reynolds number.}
\label{fig:GmaxRe}
\end{figure}

It is also interesting to explore in more detail how results depend on the Reynolds number. It is known \citep{schmid-2007} that the transient growth is a phenomenon whose dynamical importance increases with $Re$; in particular, the maximum transient growth $G_{max}$ increases as $Re^2$ for a canonical Poiseuille flow. When the flow is controlled by spanwise forcing, the $Re$ dependence is similar but not identical, as shown in \ref{fig:GmaxRe}, which neatly confirms the $Re^2$ increase for the case with $A=0$, and shows for the controlled case a slightly slower increase, fitted well by the power law $Re^{1.9}$. 

We conclude by mentioning the outcome of the same study repeated with the spanwise forcing applied to one wall only. It is confirmed that the control remains effective, but significantly less so compared to the case with forcing applied on both walls. For example, at $Re=2000$ and $A=1$, i.e. the case depicted in figure \ref{fig:Gfunc}, applying the control on both walls yields $R_{mod}=1.55$ and $R_{nmod}=3.61$, which reduce to $R_{mod}=1.14$ and $R_{nmod}=1.37$ when a sole wall is activated. This is somewhat expected, since the flow dynamics over the two walls are decoupled, while the quantities $R_{mod}$ and $R_{nmod}$ refer to the entire volume; in other words, when control is applied to one wall only, instability is nearly unaffected on the other. This observation should not be overemphasized, though, as the fact that spanwise forcing remains fully effective in a boundary for both delaying transition and decreasing the friction drag is fully assessed \cite{negi-etal-2019,skote-mishra-wu-2019}.

%% file: conclusions.tex
\section{Concluding discussion}
\label{sec:conclusions}

The present work explores the potential of a flow control technique,
originally conceived for the reduction of skin-friction drag in the
turbulent regime, to alter the linear stability characteristics of a
wall-bounded shear flow. In the simplest geometry of a plane parallel
channel, we have studied the modal and nonmodal temporal stability of
laminar, pressure-driven Poiseuille flow modified by spanwise wall
forcing. Specifically, the spanwise velocity enforced at the wall is
steady, spanwise-uniform and sinusoidally modulated along the
streamwise direction. A spanwise Stokes layer develops, the so-called
steady Stokes layer (SSL) described by Viotti \textit{et al.\ }(2009)
\cite{viotti-quadrio-luchini-2009}: owing to the convective nature of
Poiseuille flow, the steady forcing is effectively unsteady as seen by
the convecting near-wall turbulence structures, and is known to
provide large skin-friction drag reduction as well as interesting net
energy savings in the turbulent regime.

The mathematical
formulation of the stability problem is not straightforward, owing to the
streamwise-varying base flow. The sinusoidal streamwise variations of the base flow
is used to arrive at a formulation with a block-coupled system
matrix. A numerical study has investigated a large number of parameters, both
physical ($Re$, $A$, $\kappa$, $\beta$) and related to the numerical
discretization ($N$, $M$, $n_{eig}$, $\Delta t$, $T_{end}$). Overall,
a total of 11,286 cases have been computed and processed. Special care
has been taken to properly select the discretization parameters for
discretization-independent solutions, while keeping the computational
cost under control.


The main results fully support the finding that a turbulent
skin-friction drag reduction technique can be employed to improve the
stability characteristics of a laminar flow. In this respect, this
conclusion reinforces similar results recently obtained numerically
(see e.g.\ Ref.\ \cite{negi-etal-2019}) in terms of transition
delay. Looking at the asymptotic behavior of the perturbation (modal
stability), the least stable eigenvalue $\lambda_1$ has been found in
our study to increase its stability margin in comparison to its
Orr--Sommerfeld counterpart (figure \ref{fig:modal}), in a way that is
directly related to the forcing intensity. Similarly, the potential
for short-time growth of energy of small perturbations is
significantly hampered in comparison to the unforced case. The relative
stability improvements are $Re$-dependent, and across the
tested values of the Reynolds number the real part of the least stable
eigenvalue has been shown to more than double, as shown in table \ref{tab:modal}: the ratio $\Re(\lambda_1)/\Re(\lambda_1^{ref})$ increases from $1.94$ at $Re=500$ to $2.36$ at $Re=2000$. The maximum energy growth decreases by 65\% at $Re=500$ and by 72\% at $Re=2000$.


Although work is still needed to improve our understanding of the entire
picture and to approach realistic applications, especially in terms of suitable actuators which must be able to satisfy the required demanding specifications, the present results
convincingly support the claim that spanwise forcing (at least in the
stationary and spatially non-uniform case considered here) is
an effective way to improve the linear stability characteristics of
plane Poiseuille flow.